\documentclass{jfm}
\usepackage[UKenglish]{babel}

\usepackage{bm}
\usepackage{xcolor}
\usepackage{amsmath}
\usepackage{amssymb}
\usepackage{graphicx}
\usepackage{mathtools} 
\usepackage{hyperref}
\usepackage{color}

\usepackage{pdflscape}

\newcommand{\nablab}{\ensuremath{\bm{\nabla}}}
\newcommand{\Ek}{\ensuremath{\mathrm{Ek}}}
\newcommand{\Rat}{\ensuremath{\mathrm{Ra}_T}}
\newcommand{\Rac}{\ensuremath{\mathrm{Ra}_C}}
\renewcommand{\Pr}{\ensuremath{\mathrm{Pr}}} 
\newcommand{\Pm}{\ensuremath{\mathrm{Pm}}}
\newcommand{\Sc}{\ensuremath{\mathrm{Sc}}}
\newcommand{\RaTc}{\ensuremath{\mathrm{Ra}_{T,\text{crit}}}}
\newcommand{\R}{\ensuremath{\mathrm{Re}}}
\newcommand{\Rm}{\ensuremath{\mathrm{Rm}}}

\newcommand{\Ro}{\ensuremath{\mathrm{Ro}}}
\newcommand{\Nu}{\ensuremath{\mathrm{Nu}}}
\newcommand{\Sh}{\ensuremath{\mathrm{Sh}}}
\newcommand{\Le}{\ensuremath{\mathrm{Le}}}
\newcommand{\Rred}{\ensuremath{\widetilde{\mathrm{Ra}}_T}}

\begin{document}


\title{Semi-convection in rotating spherical shells: flows, layers, and dynamos}


\author{Paul Pru\v{z}ina \corresp{\email{pruzinap@univ-grenoble-alpes.fr}}, Nathana\"el Schaeffer, David C\'ebron}
\affiliation{Universit\'e Grenoble Alpes, CNRS, ISTerre, Grenoble, France}


\date{\today}


\maketitle

\begin{abstract}
	
Large regions of giant planets are thought to possess unstable thermal gradients stabilised by gradients of heavy-element composition. 
Such a fluid can develop semi-convection, a double-diffusive instability driven by the unequal molecular diffusivities of heat and composition.
While previous studies have focused mainly on local Cartesian models, we investigate semi-convection in rotating spherical shells, the geometry relevant to planetary interiors, using direct numerical simulations.
In a first nonlinear phase, the flow spontaneously forms concentric density staircases composed of well-mixed layers separated by thin, strongly stratified interfaces.
We propose scalings for both the thickness of these layers and their survival time in terms of the rotation rate and stratification.
Over longer timescales, layers merge to produce statistically steady states consisting either of a fully convective shell, or of a convective interior overlain by a persistent stably stratified layer (SSL),
depending on the balance between stratification and rotation.
Dynamo simulations show that flow within the convective region can generate a self-sustained magnetic field, which is filtered by zonal flows in the overlying SSL. This results in a strongly dipolar and axisymmetric external field, in encouraging agreement with Saturn’s magnetic field.
Across the explored parameter range, both the Rossby number and the thickness of the stably stratified layer are governed by a single combination of control parameters.
This enables identification of a regime favourable to planetary-like dynamos.
\end{abstract}
\section{Introduction}

Fluids that are convectively stable due to a bottom-heavy compositional gradient, but with a destabilising thermal gradient (i.e. stable according to the Ledoux criterion, but unstable by the Schwarzschild criterion), may undergo an instability known as semi-convection (SC) or oscillatory double-diffusive convection (ODDC). Semi-convection is one regime of double-diffusive convection, an instability caused by the effect on the density of two scalars that diffuse at different rates. For example, in gas giant planets, the thermal diffusivity can be up to $100$ times the diffusivity of heavy elements \citep{french2012ab}. The opposite regime, in which the thermal gradient is stable, but a compositional gradient is unstable, is called fingering convection. Both regimes were originally discovered in physical oceanography \citep{stern1960salt,veronis1965finite}, but the same physical process underlies instability in a planetary context, with a range of studies extending the concept in the following years \citep{kato1966overstable,stevenson1977semitheory,stevenson1977dynamics}. The transport properties of a semi-convective fluid are significantly different to one that is Schwarzschild-stable, with significantly enhanced transport of heat and composition compared to purely diffusive processes \citep{wood2013new,moll2016new}.

In astrophysics, semi-convection is of particular interest in gaseous planets such as Jupiter and Saturn \citep{muller2020challenge,stevenson2022mixing}. These planets are composed of a surface layer of molecular hydrogen, which overlays a deep core of metallic hydrogen, which has high electrical conductivity. The fluid density is affected by both the temperature and the concentration of heavy elements, most importantly the relative fraction of helium in the metallic hydrogen layer. Recent structure models predict the existence of wide regions with destabilising temperature gradients, but stabilising compositional gradients, that result in a convectively stable state. \citet{debras2019new} showed that to reproduce observations from the Juno and Galileo missions, the internal structure of Jupiter must contain such a stably stratified layer between two convective envelopes, between about $80$ and $90\%$ of the planetary radius. To match the thermal properties of the convective zones, a destabilising temperature gradient must exist in this region, providing the necessary conditions for semi-convection. Likewise, \cite{leconte2012new,leconte2013layered} suggest that a region of semi-convection could explain Saturn's excessively high luminosity. \cite{mankovich2021diffuse} proposed an interior model of Saturn consistent with gravity and seismic data, with a stably stratified region extending to $60\%$ of the planetary radius.  Similar semi-convective regions are also predicted at the edge of the convective cores of intermediate and high-mass stars \citep[e.g.][]{schwarzschild1958evolution,merryfield1995hydrodynamics}. Semi-convection also occurs in terrestrial polar oceans (where it is called `diffusive convection'), with the stabilising compositional gradient provided by salinity \citep[e.g.][]{timmermans2008eddies}. 

Semi-convection is sometimes referred to in astrophysical literature as `layered convection', referring to the tendency of semi-convective fluids to spontaneously form `staircases' consisting of well-mixed layers separated by sharp density jumps. This phenomenon is well-documented in periodic box simulations \citep[e.g.][]{rosenblum2011turbulent,mirouh2012new}, and more recently in full sphere simulations \citep{fuentes20253d}. Layer formation is due to the so-called `$\gamma$-instability' \citep{radko2003mechanism}, where competing influences of thermal and compositional fluxes lead to a secondary instability on the semi-convective state. \citet{pruzina2025onedimensional} proposed a one-dimensional model of this layering process for parameters relevant to planetary applications.

The effect of rotation on layer formation was studied in 3D periodic boxes by \citet{moll2017effect}. They showed that sufficiently fast rotation rates affect even the small-scale double-diffusive instability, preventing layer formation.
Interestingly, \citet{fuentes20253d} showed that a more moderate rotation rate can prolong the lifespan of a staircase, although in all cases the layers eventually coarsen through mergers until a single convective region fills the domain.
This gradual coarsening is widely documented across numerical studies of layering \citep{radko2003mechanism,mirouh2012new,pruzina2025onedimensional}, following a pattern described by \citet{radko2007mechanics}. However, such merging has not yet been convincingly demonstrated in observational studies. So far, the majority of work on astrophysical semi-convection has focused on the layered regime \citep[e.g.][]{moll2017effect,fuentes2024evolution,fuentes20253d}, leaving behaviour outside layered states relatively unexplored.

A particular feature of interest is whether semi-convection can support planetary magnetic fields. Previous work has generally assumed that planetary dynamos are generated by convective motions due to an overall statically unstable density gradient \citep[e.g.][]{christensen2006scaling,jones2014dynamo,yadav2022global}, but recent work predicting semi-convective regions \citep[e.g.][]{leconte2013layered,vazan2016evolution,debras2019new} raises the question of whether semi-convection-driven dynamos are possible. There are few prior studies on this subject, with \citet{mather2021regimes} concluding that while semi-convection dynamos in spherical shells are possible, they require parameter values far from true astrophysical conditions. \citet{pruzina2025planetary} demonstrated that simulations of semi-convection can indeed support dynamo action at low magnetic Prandtl number, showing a simulation where a deep convective dynamo is overlain by a stably stratified layer, producing a rather dipolar magnetic field on the surface.  However, this was only a preliminary demonstration, and the detailed behaviour of semi-convection dynamos remains an open question.

The related problem of rotating fingering convection was studied  in a full sphere by \citet{monville2019rotating}. They found that, compared with pure compositional convection, fingering convection increased the parameter range for linear instability, leading to large-scale convective motions and strong zonal jets. \citet{gray2025influence} have recently performed a more detailed study, finding a wide range of flows including clustering of fingers, toroidal gyres, hemispherical convection, and zonal flows, depending on the relative strengths of the thermal stratification and rotation. It seems likely that a similarly rich variety of dynamical structures may be possible in the semi-convection regime.

To model the semi-convective layer of gaseous planets, we investigate semi-convection in a spherical shell, with fixed temperature and composition at the boundaries to represent steady convecting layers above and below.
Non-linear simulations pass through a transient layered phase, eventually saturating in a state either dominated by strong zonal jets, with a wide stably stratified layer overlaying an inner convective region, or a structure where a single convective zone fills the entire domain, except for narrow thermal boundary layers.  We find that the layering instability conforms well to the predictions of the $\gamma$-instability \citep{radko2003mechanism}, but we also confirm the comment of \citet{fuentes2024evolution} that the size of the domain can constrain the development of layers. We investigate the onset of dynamo action and discuss the effect of different flow structures on the magnetic field. We find that a dipolar structure is favoured in the jet-dominated flow regime with a wide SSL, where a magnetic field is generated in the inner convective layer, but decays as a function of radius in the SSL. The dipolar component decays the slowest, resulting in a much more dipolar surface field than in the interior. On this basis, we identify a region of parameter space that is most favourable to producing planetary-like magnetic fields, and compare simulations in this region to Saturn's magnetic field.

The paper is structured as follows. In \S~\ref{sec:eqns} we introduce the equations of thermo-compositional convection and discuss the non-dimensional parameters governing the flow. In \S~\ref{sec:stability} we review the hydrodynamic stability of the system, before presenting nonlinear simulations in \S~\ref{sec:non-lineardynamics}, discussing separately the layered phase of evolution, the saturated states, and the different flow regimes that develop. Finally, in \S~\ref{sec:dynamo}, we present the results of magnetohydrodynamic simulations, and propose a criterion for producing planetary-like magnetic fields in simulations.

\section{Governing equations of thermocompositional convection}\label{sec:eqns}

\begin{figure}
    \centering
    \includegraphics[width=0.4\textwidth]{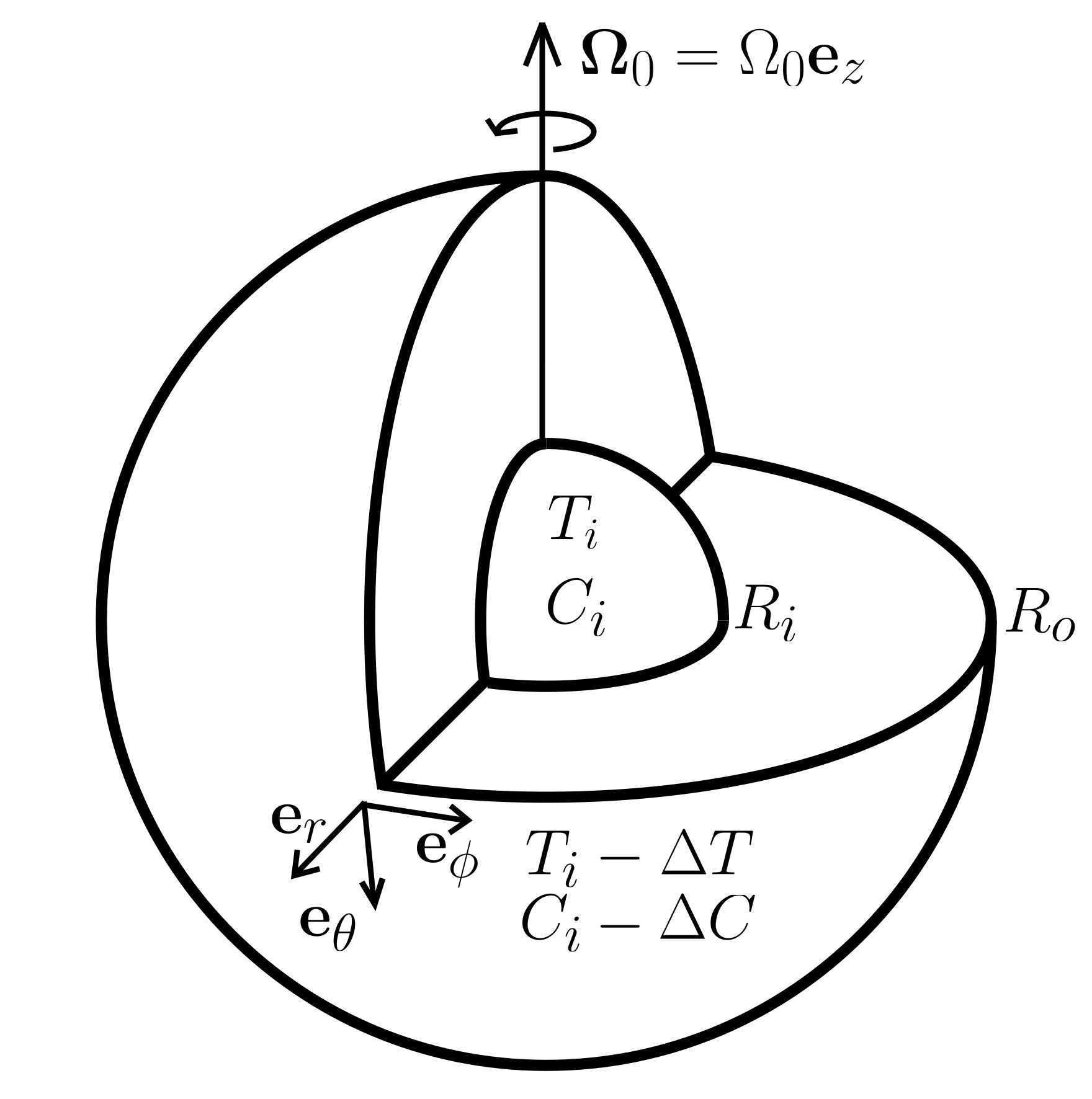}
    \caption{Sketch of the physical setup for the problem.}
    \label{fig:rotatingsphere}
\end{figure}

\begin{table}
    \centering
    \begin{tabular}{c|c|}
         Symbol&Name\\
         $g$&Gravitational acceleration\\
        $\rho_0$&Reference density\\
        $\nu$&Viscous diffusivity\\
        $\mu_0$&Vacuum magnetic permeability\\
        $\kappa_T$&Thermal diffusivity\\
        $\kappa_C$&Compositional diffusivity\\
        $\eta$&Magnetic diffusivity\\
        $\alpha_T$&Thermal expansion coefficient\\
        $\alpha_C$&Compositional contraction coefficient\\      
        $R_o$&Outer radius\\
        $R_i$&Inner radius\\
        $\Omega$&Rotation rate\\
        $\Delta T$&Temperature difference\\
        $\Delta C$&Composition difference\\
    \end{tabular}
    \begin{tabular}{c|c|c|c|}
    	Symbol&Name&Definition&Range in simulations\\
         $\Ek$&Ekman number&$\nu/(R_o^2\Omega)$&$10^{-6}$ -- $10^{-3}$\\
         $\Pr$&Prandtl number&$\nu/\kappa_T$&$0.01$ -- $3$\\
         $\Sc$&Schmidt number&$\nu/\kappa_C$&$0.1$ -- $30$\\
         $\Le$&Lewis number&$\Sc/\Pr$&$10$\\
         $\Pm$&Magnetic Prandtl number&$\nu/\eta$&$0.1$ -- $1$\\
         $\Delta R$&Shell thickness&$(R_o-R_i)/R_o$&$0.3$ -- $0.8$\\
         $\Rat$&Thermal Rayleigh number&$\alpha_Tg\Delta T R_o^3/(\kappa_T\nu)$&$10^5$ -- $5\times10^9$\\
         $\Rac$&Compositional Rayleigh number&$\alpha_Cg\Delta C R_o^3/(\kappa_C\nu)$&$10^6$ -- $6\times10^{10}$\\
         $R_\rho$&Density ratio&$|\Rac|/(|\Rat|\Le)$&$1$ -- $3.3$\\
         $\Rred$&Reduced Rayleigh number&$\Ek^{4/3}\Rat$&$2$ -- $4000$\\
         $\R$&Reynolds number&$ UL/\nu$&$5$ -- $1.4\times10^5$\\
         $\Rm$&Magnetic Reynolds number&$UL/\eta$&$40$ -- $2200$\\
         $\Ro$&Rossby number&$\Ek\R$&$5\times10^{-5}$ -- $2$
    \end{tabular}
    \caption{Dimensional and dimensionless parameters governing the dynamics of rotating semi-convection.}
    \label{tab:parameters}
\end{table}
We consider a fluid in a spherical shell of inner radius $R_i$ and outer radius $R_o$, rotating at a rate $\bm{\Omega_0}=\Omega_0\bm{e}_z$, as shown in fig.~\ref{fig:rotatingsphere}. We use spherical polar coordinates, with radial, polar and azimuthal basis vectors $\{\bm{e}_r,\bm{e}_\theta,\bm{e}_\phi\}$ and position vector $\bm{r}$ such that $\theta=0$ for $\bm{r}=\bm{e}_z$. The fluid velocity $\bm{u} = u_r\bm{e}_r + u_\theta \bm{e}_\theta + u_\phi\bm{e}_\phi$  evolves according to the Navier-Stokes equations. We adopt the Boussinesq approximation, which neglects differences in density except in terms multiplied by gravity. The magnetic field $\bm{b}$ evolves according to the induction equation, and affects the velocity through the Lorentz force. We adopt a linearised version of the equation of state for the density $\rho$
\begin{equation}
    \rho = \rho_0(1-\alpha_T(T_0+T)+\alpha_C(C_0+C)),\label{eqn:eos}
\end{equation}
where $\rho_0$ is a reference density, $T_0$ and $C_0$ are background temperature and composition fields, $T$ and $C$ are the perturbation temperature and composition, and $\alpha_T$ and $\alpha_C$ are the coefficients of thermal expansion and compositional contraction. The temperature and composition obey advection--diffusion equations in the presence of the background fields. We assume a constant gravitational acceleration in the radial direction $\bm{g}=-g\bm{e}_r$, giving a radial buoyancy force of $g(\rho_0-\rho(\bm{x},t)) \bm{e}_r$. In gas giants, this is not an unreasonable assumption: for Saturn, \cite{movshovitz2020saturn} give a difference of only $20\%$ between $g(0.1R_o)$ and $g(0.7R_o)$. We write the thermal and compositional differences across the depth as $\Delta T = T_0(R_i)-T_0(R_o)$ and $\Delta C = C_0(R_i)-C_0(R_o)$, respectively.

The dynamics depend on a number of physical parameters, which are defined in Table~\ref{tab:parameters}. Using these parameters, and with a length-scale based on the sphere's outer radius $R_o$ and the viscous diffusion timescale $R_o^2/\nu$, we adopt dimensionless variables
\begin{equation}
\begin{split}
    &\bm{x} = \frac{\hat{\bm{x}}}{R_o},\qquad t = \frac{\nu}{R_o^2}\hat{t},\qquad \bm{u} = \frac{R_o}{\nu}\hat{\bm{u}},\\
    &T = \frac{g\alpha_TR_o^3}{\nu^2}\hat{T},\qquad C = \frac{g\alpha_C R_o^3}{\nu^2}\hat{C},\qquad \bm{b}=\frac{R_o}{\nu\sqrt{\rho_0\mu_0}}\hat{\bm{b}},
\end{split}
\end{equation}
where hats denote dimensional quantities. The dynamics are governed by the non-dimensional system of equations
\begin{eqnarray}
    \frac{\partial}{\partial t}\bm{u} + \left(\frac{2}{\Ek}\bm{e}_z+\nablab\times\bm{u}\right)\times\bm{u} &=& -\nablab p +\left(T-C\right)\bm{e}_r + \left(\nablab\times\bm{b}\right)\times\bm{b} + \nabla^2\bm{u},\label{eqn:momentumequation}\\
    \frac{\partial}{\partial t}T + ( \bm{u}\cdot \nablab)\left(  T+ T_0\right) &=& \frac{1}{\Pr} \nabla^2  T,\label{eqn:Tequation}\\
    \frac{\partial}{\partial t} C + ( \bm{u}\cdot \nablab)\left( C+ C_0\right) &=& \frac{1}{\Sc} \nabla^2 C,\label{eqn:Cequation}\\
    \frac{\partial}{\partial  t} \bm{b} &=&  \nablab\times\left( \bm{u}\times \bm{b}-\frac{1}{\Pm} \nablab\times \bm{b}\right),\label{eqn:bequation}
\end{eqnarray}
together with the solenoidal constraints $\nablab\cdot \bm{u} = \nablab\cdot \bm{b} = 0$. The dimensionless numbers $\Ek$, $\Pr$, $\Sc$ and $\Pm$ are defined in Table~\ref{tab:parameters}. For the majority of this work, we shall consider only the hydrodynamic problem (i.e. with $\bm{b}\equiv\bm{0}$); we return to the full magneto-hydrodynamic system in \S~\ref{sec:dynamo}.

A steady basic state exists with $\bm{u}= \bm{0}$, and background temperature and salinity fields given by
\begin{equation}
    T_0(r) = \frac{\Rat}{\Pr}\delta(r),\quad C_0(r) = \frac{\Rac}{\Sc}\delta(r),\label{eqn:T0C0}
\end{equation}
where the thermal and compositional Rayleigh numbers are defined in Table~\ref{tab:parameters}, and $\delta(r)$ is the spherically symmetric solution to Laplace's equation $\nabla^2\delta=0$ in a spherical shell:
\begin{equation}
    \delta(r) = \frac{1-\Delta R}{\Delta R}\frac{1-r}{r},\label{eqn:delta(r)}
\end{equation}
such that $\delta(1-\Delta R) = 1$ and $\delta(1) = 0$.

To approximate the effect of convective regions above and below the domain, we fix the temperature and composition on the boundaries. The different temperatures and compositions above and below are accounted for by the background fields $T_0$ and $C_0$, so the perturbation fields are set to zero. The velocity obeys stress free, no penetration conditions on the boundaries:
\begin{equation}
    T =C = \frac{\partial}{\partial r}\left(\frac{u_\theta}{r}\right) =  \frac{\partial}{\partial r}\left(\frac{u_\phi}{r}\right)= u_r= 0\text{ at } r = 1-\Delta R, 1.\label{eqn:bcs}
\end{equation}
For the magnetic field, we adopt an insulating boundary condition on the outer sphere. In practice, this means that the toroidal component of $\bm{b}$ is set to zero, and the poloidal component matches a potential field outside \citep[see][]{christensen2015numerical}. We assume that the volume within the inner sphere is electrically conducting, as suggested by \citet{debras2019new}, so solve \eqref{eqn:bequation} in this region with $\bm{u}(r<R_i)=\bm{0}$.  Both $\bm{b}$ and $\partial\bm{b}/\partial r$ are continuous at both boundaries.

\section{Hydrodynamic stability}\label{sec:stability}
In this section, we present a number of linear stability results relevant to semi-convection. We detail the classical stability properties resulting from a local analysis in a plane layer geometry, before discussing the stability in rotating spherical shells. Further details of the latter are provided in Appendix~\ref{sec:linearonset}. Finally, we discuss the criteria leading to the formation of density layers.

The density gradient can be used to define the Brunt-V\"ais\"al\"a frequency $N$, defined as the frequency at which a vertically displaced particle will oscillate in the stratified environment. In dimensional form, $N^2$ is given by
\begin{equation}
    N^2 = -g\bm{e}_r\cdot\nablab \left( \frac{\rho}{\rho_0} \right).
\end{equation}
A generalisation can be made to split the background frequency $N_0^2=N^2(T=C=0)$ into contributions from temperature and composition such that $N_0^2 = N_{0,T}^2 + N_{0,C}^2$ as follows
\begin{equation}
    N_{0,T}^2 = \alpha_Tg\bm{e}_r\cdot\nablab T_0,\quad N_{0,C}^2 = -\alpha_Cg\bm{e}_r\cdot\nablab C_0.
\end{equation}
In dimensionless form, with the background states given by \eqref{eqn:T0C0}, these are given by 
\begin{equation}
    N_{0,T}^2 = -\frac{\Rat}{\Pr}\frac{1-\Delta R}{\Delta R}\frac{1}{r^2} ,\quad N_{0,C}^2 = +\frac{\Rac}{\Sc}\frac{1-\Delta R}{\Delta R}\frac{1}{r^2}.
\end{equation}

A number of standard linear stability results may be derived from a plane layer geometry, with uniform background temperature and compositions gradients (i.e. $N^2_{T,C}$ constant).
The Schwarzschild criterion \citep{schwarzschild1958structure} states that a fluid is stable to thermal convection if 
\begin{equation}
    N_{0,T}^2 > 0.
\end{equation}
The related Ledoux criterion \citep{ledoux1947stellar}, which includes the effects of composition, states that a fluid is stable to convective motions if 
\begin{equation}
    N_0^2 > 0.\label{eqn:ledouxcriterion}
\end{equation}
The regime of interest for semi-convection is where fluid is Ledoux-stable, but Schwarzschild unstable, i.e.
\begin{equation}
    N_0^2>0 \quad\mathrm{and}\quad N_{0,T}^2<0,
\end{equation}
or, in terms of the Rayleigh numbers,
\begin{equation}
    \Rat, \Rac > 0.
\end{equation}
Despite being stable to convection according to the Ledoux criterion \eqref{eqn:ledouxcriterion}, \citet{walin1964note} and \citet{kato1966overstable} demonstrated that a semi-convective instability in a plane layer can occur if 
\begin{equation}
    \Rat > \Rat^{SC} = \Pr\left(\frac{\Pr+1/\Le}{\Pr+1}\frac{\Rac}{\Sc} + \left(1+\frac{1}{\Le}\right)\left(1+\frac{1}{\Le\Pr}\right)\frac{27\pi^4}{4}\right),\label{eqn:semiconvRa}
\end{equation}
where $\Le=\kappa_T/\kappa_C = \Sc/\Pr$ is the Lewis number or diffusivity ratio. For sufficiently high Rayleigh numbers, the second term may be neglected. 

The range of double-diffusive instability is commonly discussed in terms of the density ratio
\begin{equation}
    R_\rho = \frac{\alpha_C\left|\nablab C_0\right|}{\alpha_T\left|\nablab T_0\right|} = \frac{1}{\Le}\frac{\Rac}{\Rat}.\label{eqn:Rrhodef}
\end{equation}
The quantity $R_\rho$ is defined by \eqref{eqn:Rrhodef} for semi-convection, but by its inverse in the fingering regime, such that $R_\rho$ is always greater than unity in the double-diffusive regime of interest. 

Rewriting \eqref{eqn:semiconvRa} in terms of $R_\rho$, semi-convection occurs in the range 
\begin{equation}
    1< R_\rho<\frac{\Pr+1}{\Pr+1/\Le},\label{eqn:semi-convectionrange}
\end{equation}
with the fluid statically stable for larger values of $R_\rho$, and convectively unstable by \eqref{eqn:ledouxcriterion} for smaller values. Note that if $\Pr$ is large, then the range of $R_\rho$ for semi-convection becomes very narrow; this is the limit where viscous diffusion is significantly faster than thermal diffusion. In astrophysical fluids, $\Pr$ is generally much less than unity, so semi-convection can occur across a wide range of density ratios.

Including the Coriolis force adds significant complexity. \citet{pearlstein1981effect} showed that the effect of rotation depends strongly on the size of the Prandtl and Schmidt numbers, and can exert either a stabilising or destabilising effect. In the case where $\Pr<1<\Sc$ \citep[as suggested by][]{french2012ab}, there is a range of parameter space in which increasing $\Rac$ (that is, increasing the stabilising gradient) can in fact decrease the critical value of $\Rat$ for instability. However, this effect is very small, with a change of $\RaTc$ of less than $1\%$. More relevant to the spherical geometry considered here is the analytical stability analysis of \citet{busse2002low} in a cylindrical annulus with sloping top and bottom boundaries \citep[later corrected by][]{monville2019rotating}. 
While these papers focused mainly on the fingering regime, this analytical approach can equally be applied to semi-convection. In an appendix dedicated to SC, \citet{monville2019rotating} demonstrated a reduction in $\RaTc$ of an order of magnitude compared to that of rotating thermal convection.

A linear analysis of the onset of semi-convection in the setup of the present study is reported in Appendix A.
The onset of instability can be summarised to occur when $\Rat$ exceeds both the critical Rayleigh number for rotating thermal convection $\Rat^{rot}\sim\Ek^{4/3}$ \citep{dormy2004onset}, and the critical value for non-rotating semi-convection \eqref{eqn:semiconvRa}. In this paper, we focus on the regime where $\Rat^{SC}>\Rat^{rot}$, which occurs when $\Rac\gtrsim\Le\Ek^{-4/3}$. We note also the existence of a `tongue' of instability that exists for $\Rat<\Rat^{rot}$ in some cases, which is not the focus of this study.

The formation of staircases is due to a secondary instability which acts on the fields resulting from the double-diffusive instability. The $\gamma$-instability theory \citep{radko2003mechanism} (developed for the fingering regime) states that if $F_T$ and $F_C$ are the turbulent temperature and compositional fluxes, and $\gamma=F_C/F_T$ is their ratio, then layers will form if 
\begin{equation}
    \frac{\partial\gamma}{\partial R_\rho} < 0.\label{eqn:gammainstability}
\end{equation}
It was shown by \citet{rosenblum2011turbulent,mirouh2012new} that the same condition is valid in the semi-convection regime.
Similarly to $R_\rho$, $\gamma$ is generally defined as its reciprocal in the fingering regime, so condition \eqref{eqn:gammainstability} is unchanged.

In practice, \eqref{eqn:gammainstability} is satisfied in a range 
\begin{equation}
    1\leq R_\rho\leq R_{\rho,\text{layer}}<\frac{\Pr+1}{\Pr+1/\Le},
\end{equation}
where the specific value of $R_{\rho,\text{layer}}$ depends on the parameters of the fluid \citep[e.g.][]{rosenblum2011turbulent,mirouh2012new}. However, condition~\eqref{eqn:gammainstability} requires $\partial\gamma/\partial R_\rho$ to be calculated across a range of simulations for different values of $R_\rho$, so on its own it cannot be used as a predictor of layering a priori. \citet{mirouh2012new} and \citet{pruzina2025onedimensional} have both proposed empirically-derived parametrisations  to model $\gamma$ in terms of $R_\rho$, $\Pr$ and $\Le$.

\section{Dynamics in the non-linear regime}\label{sec:non-lineardynamics}

We now investigate the dynamics beyond the onset of instability, using the XSHELLS code \citep{schaeffer2013efficient,monville2019rotating} to solve the hydrodynamic Boussinesq equations \eqref{eqn:momentumequation}--\eqref{eqn:Cequation} (with $\bm{b}=\bm{0}$) in a spherical shell  with boundary conditions \eqref{eqn:bcs}. 

Throughout this work, unless otherwise specified, we take $\Pr=0.3$ and $\Sc = 3$, giving a Lewis number of $10$. These values are commonly used for simulations of astrophysical fluids, as they allow the low-$\Pr$ regime to be captured, while remaining numerically accessible at relatively low resolutions \citep[e.g.][]{breuer2010thermochemically}. We take $\Delta R=0.5$ as an intermediate size for the shell, noting that a thinner shell would be more computationally demanding.

\begin{figure}
	\centering
	\includegraphics[width=\textwidth]{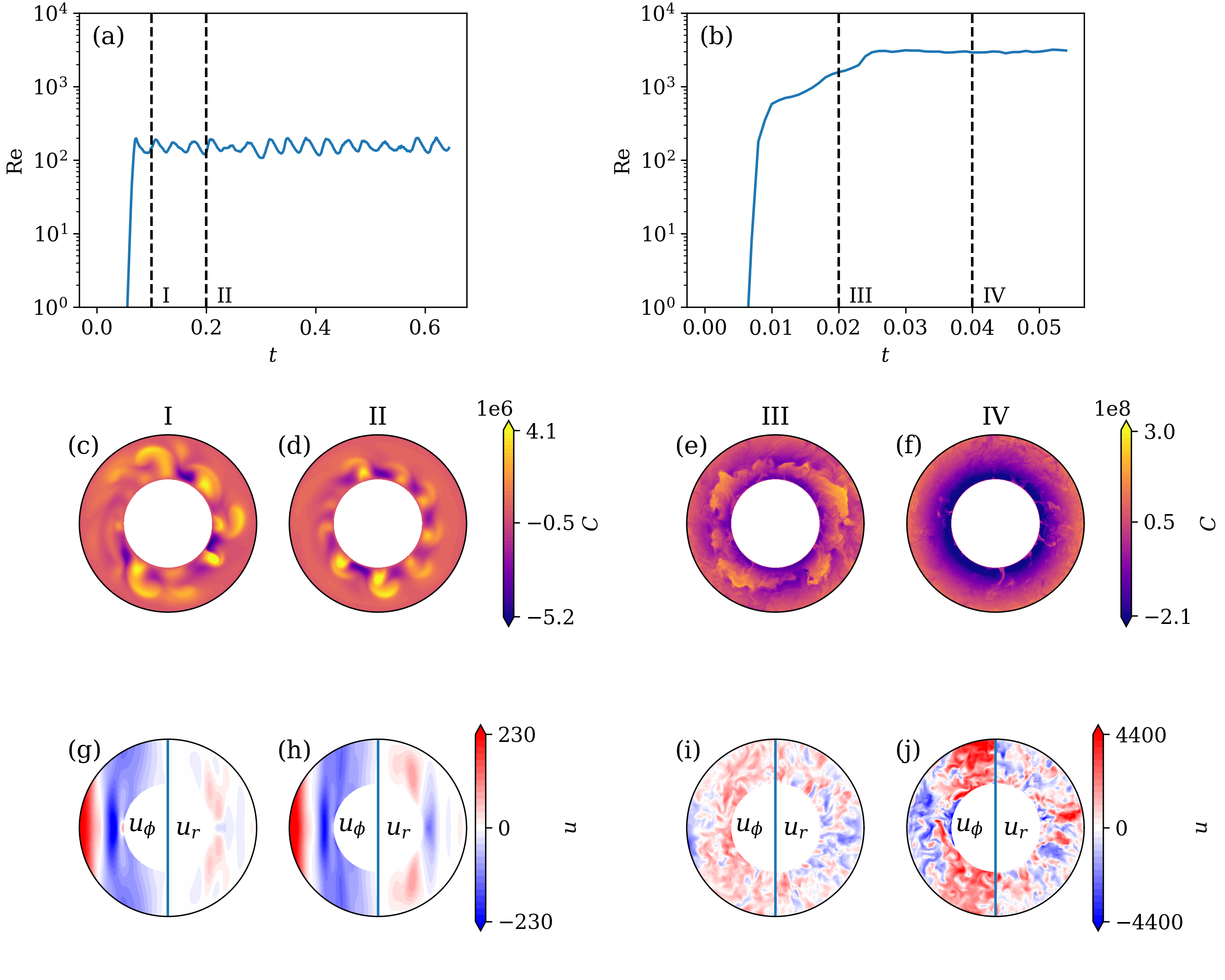}
	\caption{Results from two XSHELLS simulations with $\Ek=10^{-4}$, $\Pr=0.3$, $\Sc=3$, $\Delta R=0.5$, $R_\rho=1.2$. Left-hand columns take $\Rat=3.52\times10^6$, right-hand columns $\Rat=1.11\times10^8$. (a)--(b) Time series of the Reynolds number $\R$. (c)--(f) show equatorial snapshots of the  composition perturbation $C$ at times marked with dashed lines on (a) and (b), and labelled ``I''--``IV''. (g)--(j) show meridional snapshots of the radial and azimuthal velocities $u_r$ and $u_\phi$, at the same times. }
	\label{fig:firstsnapshots}
\end{figure}

To motivate what follows, we first show the results of two contrasting simulations in fig.~\ref{fig:firstsnapshots}, for $\Ek=10^{-4}$, $\Pr=0.3$, $\Sc=3$ and $R_\rho=1.2$. We show one simulation with weak forcing ($\Rat=3.52\times10^6$), and another with stronger forcing ($\Rat=1.11\times10^8$). In each case, the system is initialised with a small amplitude random perturbation to the velocity field $\bm{u}$. We characterise the flow strength using the Reynolds number
\begin{equation}
    \R(t) = \frac{\langle u\rangle_\text{rms}R_o}{\nu},
\end{equation}
where $\langle u\rangle_\text{rms} = \left(\int_V \bm{u}\cdot\bm{u}\mathrm{d}\\
V/V\right)^{1/2}$ is the root-mean-squared velocity (with $V$ the volume of the domain).

Fig.~\ref{fig:firstsnapshots}(a)--(b) show how the Reynolds number evolves for each simulation, while figs.~\ref{fig:firstsnapshots}(c)--(j) show equatorial snapshots of the composition perturbation $C$, and meridional snapshots of the radial and azimuthal velocities $u_r$ and $u_\phi$, at two times for each simulation. For the weak forcing case, fig.~\ref{fig:firstsnapshots}(a) shows that $\R$ first increases exponentially, then saturates into quasi-periodic oscillations of amplitude $\sim50$ around a mean value $\R\approx 150$, with statistically stable convective motions continuing for a long time. Figs~\ref{fig:firstsnapshots}(c)--(d) show no significant difference between the composition fields at the two different times shown, with a wave-like response throughout. Figs~\ref{fig:firstsnapshots}(g)--(h) show that the velocity field at both times is dominated by a columnar zonal jet structure, with relatively weaker radial motions.

The behaviour of the simulation with $\Rat=1.11\times10^8$ is different. Fig.~\ref{fig:firstsnapshots}(b) shows that, after an initial exponential growth phase, $\R$ appears to level off, before increasing again twice, eventually saturating at $\R\approx2000$. The composition snapshot during the intermediate plateau (labelled ``III'') in fig.~\ref{fig:firstsnapshots}(e)  show a two-layer structure, with two regions within which $C$ tends to increase with radius, separated by a rather sharp interface, in the form common to double-diffusive layers \citep[see, e.g.][]{fuentes2024evolution}. Later on, in the snapshots labelled ``IV'' (fig.~\ref{fig:firstsnapshots}(f)), the two layers have merged, and the radial gradient of $C$ is relatively uniform. The radial and zonal velocities shown in fig.~\ref{fig:firstsnapshots}(i)--(j) have similar magnitudes, and the columnar flow structure is much less evident than in the $\Rat=3.52\times10^6$ simulation. 

To characterise the flows observed in our simulations, we define a number of dimensionless parameters. 
Turbulent transport of heat and composition throughout the domain is quantified by the Nusselt and Sherwood numbers, defined as the ratio of the total radial scalar transport to purely conductive transport. These are defined at each radius as 
\begin{eqnarray}
    \Nu(r,t) &=& \frac{-\langle u_r\left(T+T_0\right)\rangle_S + \Pr^{-1}\langle\partial (T+T_0)/\partial r\rangle_S}{\Pr^{-1}\langle\partial T_0/\partial r\rangle_S},\label{eqn:Nusselt}\\
    \Sh(r,t) &=& \frac{-\langle u_r\left(C+C_0\right)\rangle_S + \Sc^{-1}\langle\partial (C+C_0)/\partial r\rangle_S}{\Sc^{-1}\langle\partial C_0/\partial r\rangle_S},\label{eqn:Schmidt}
\end{eqnarray}
where $\langle\cdot\rangle_S$ represents the mean across a surface of constant radius. In a purely conductive state, $\Nu=\Sh=1$, the values increase as the intensity of the convection increases.
In practice, we evaluate $\Nu$ and $\Sh$ at the inner or outer boundary, where the convective fluxes vanish due to the no-penetration condition on the radial velocity:
\begin{equation}
    \Nu(t) = \left[\frac{\langle\partial (T+T_0)/\partial r\rangle_S}{\langle\partial T_0/\partial r\rangle_S}\right]_{r=1-\Delta R,1},\quad \Sh(t) = \left[\frac{\langle\partial (C+C_0)/\partial r\rangle_S}{\langle\partial C_0/\partial r\rangle_S}\right]_{r=1-\Delta R,1}.
\end{equation}

Also of interest is the convective power, defined as the work done by buoyancy forces, with thermal and compositional components given in dimensionless form by
\begin{equation}
    P_T(t) = \langle\left(T+T_0\right)u_r\rangle_V,\quad P_C(t) = -\langle\left(C+C_0\right)u_r\rangle_V,
\end{equation}
with $\langle\cdot\rangle_V$ denoting the volume average over the whole domain. In the semi-convection regime, motions are driven by a positive $P_T$, while $P_C$ is negative, opposing fluid motions. Where convection occurs, $P=P_T+P_C$ is positive. 

In general, $\R$, $\Nu$, $\Sh$, $P_T$ and $P_C$ are time-varying quantities, but it is more useful to consider their time average, denoted by an overbar. For some time-dependent quantity $q$, We define
\begin{equation}
    \overline{q} = \frac{1}{t_2-t_2}\int_{t_1}^{t_2} q(t) dt, \label{eqn:timeave}
\end{equation}
where $t_1$ is generally the time at which the simulation reaches a statistically steady state, and $t_2$ is the end of the simulation.

\subsection{The layered phase of evolution}
In this section, we study the transient layered phase in more detail, quantifying the dependence of the layers on the system parameters and considering their long-term fate. Although double-diffusive layers are well-documented in periodic boxes \citep[e.g.][]{rosenblum2011turbulent,mirouh2012new,wood2013new}, significantly less work has been done in a spherical geometry. According to the widely accepted theory of \citet{radko2003mechanism}, the layering instability depends only on the density ratio $R_\rho$, and a semi-convective fluid will develop density layers if the ratio of the composition to temperature fluxes $\gamma=F_C/F_T$ decreases as a function of the density ratio. This theory has been confirmed by numerous works on simulations in both fingering and semi-convection regimes \citep[e.g.][]{radko2003mechanism,wood2013new}. However, this is a linear theory, based on local uniform gradient background states, without the influence of boundaries. The simulations shown in figures~\ref{fig:firstsnapshots} use the same value of $R_\rho$, but do not both produce layers, showing that in the bounded geometry that we consider, layering must also depend on the stratification or forcing.

Recently, \citet{fuentes20253d} conducted simulations of semi-convection in full spheres, finding that the semi-convection instability developed into a series of convective layers, which were gradually eroded by a strong outer layer growing into the interior. In these simulations, faster rotation rates resulted in longer-lasting layers, with the outer convective layer growing towards the centre more slowly. 
This runs contrary to the findings of \citet{moll2017effect}, who demonstrated that fast rotation suppresses layer formation. If the modified Taylor number
\begin{equation}
	Ta^*=\frac{4\Pr^2}{\Ek^2\Rat}\gg1,\label{eqn:Ta*}
\end{equation}	
then the rotation is so strong that it affects even the small-scale double-diffusive instability. \citet{fuentes2024evolution} commented that this layer suppression was due to the periodic box simulations of \citet{moll2017effect} using a domain too small for layers to form. Such a domain size constraint has been documented by \citet{mirouh2012new} and \citet{wood2013new}, who found that a domain of depth $\gtrsim 50d$ was necessary for layer formation, where 
\begin{equation}
	d=\left(\frac{\kappa_T\nu}{\alpha_Tg\left(\partial T_0/\partial r\right)}\right)^{1/4},\label{ref:locallength}
\end{equation}	
represents the characteristic scale for the semi-convection instability.
By redefining the Rayleigh number in terms of the thermal gradient, rather than temperature difference,
\begin{equation}
	\Rat=\frac{\alpha_Tg\partial T_0/\partial r R_o^4}{\kappa_T\nu} = \frac{N^2_{0,T}R_0^4}{\kappa_T\nu},\label{eqn:gradientRayleigh}
\end{equation}
we note that $\Rat^{1/4}(\Delta R/R_o)$ provides a measure of the shell thickness normalised by $d$; hence the condition of \citet{mirouh2012new,wood2013new} translates into a requirement that $\Rat>(50R_o/\Delta R)^4$ for layer formation. 
While a domain-size constraint is a potential reason for layers to be suppressed at lower $Ta^*$, the high-$Ta^*$ simulations of \citet{moll2017effect} suppress the $\gamma$-instability itself.

%

To study the effects of both parameters, we discuss simulations along two transects: with $\Rac$ fixed while $R_\rho$ varies, and vice versa. In both transects, the strength of the forcing ($\Rat$) varies; however, the fixed-$R_\rho$ transect runs approximately parallel to the onset of semi-convection, while the fixed-$\Rac$ transect spans the width from the onset of convection ($R_\rho=1$) to past the onset of semi-convection $R_\rho=(\Pr+1)/(\Pr+1/\Le)$

\subsubsection{Effect of $R_\rho$ on layer formation}\label{sec:layerRp}
To measure the effect of $R_\rho$ on layering, we show simulations with $\Rac=2.33\times10^8$ fixed, and $1\leq R_\rho\leq3.3$, for $\Pr=0.3$, $\Sc=3$, $\Delta R=0.5$, $\Ek=10^{-4}$. This transect runs across the semi-convection region, from the onset of top-heavy convection, to the stable region. From the temperature and composition equations \eqref{eqn:Tequation}--\eqref{eqn:Cequation}, and the definitions of the background fields \eqref{eqn:T0C0}, we define the local radial temperature and composition fluxes as 
\begin{eqnarray}
     F_T^{local} &=& \frac{1}{\Pr}\frac{\partial}{\partial r}\left(T_0+T\right) + \left(T+T_0\right)u_r,\\
     F_C^{local} &=& \frac{1}{\Sc}\frac{\partial}{\partial r}\left(C_0+C\right) + \left(C+C_0\right)u_r.
\end{eqnarray}
Upon averaging over a spherical surface, comparison with \eqref{eqn:Nusselt}--\eqref{eqn:Schmidt} shows that
\begin{eqnarray}
	\langle F_T^{local}\rangle_S = \frac{\langle\partial T_0/\partial r\rangle_S}{\Pr} \Nu(r,t),\\
	\langle F_C^{local}\rangle_S = \frac{\langle\partial C_0/\partial r\rangle_S}{\Sc} \Sh(r,t).\\
\end{eqnarray}
Hence the flux ratio may be written as 
\begin{equation}
	\gamma=\frac{F_C}{F_T}=\frac{\Sh}{\Nu}\frac{R_\rho}{\Le}\label{eqn:gamma}
\end{equation}
In a pure conductive state (i.e. $\bm{u}=\bm{0}$), the Nusselt and Sherwood numbers are equal to unity, and hence $\gamma=R_\rho/\Le$. Recall from \S~\ref{sec:stability} that layering is predicted by the $\gamma$-instability where $\partial \gamma/\partial R_\rho<0$.

\begin{figure}
    \centering
    \includegraphics[width=\textwidth]{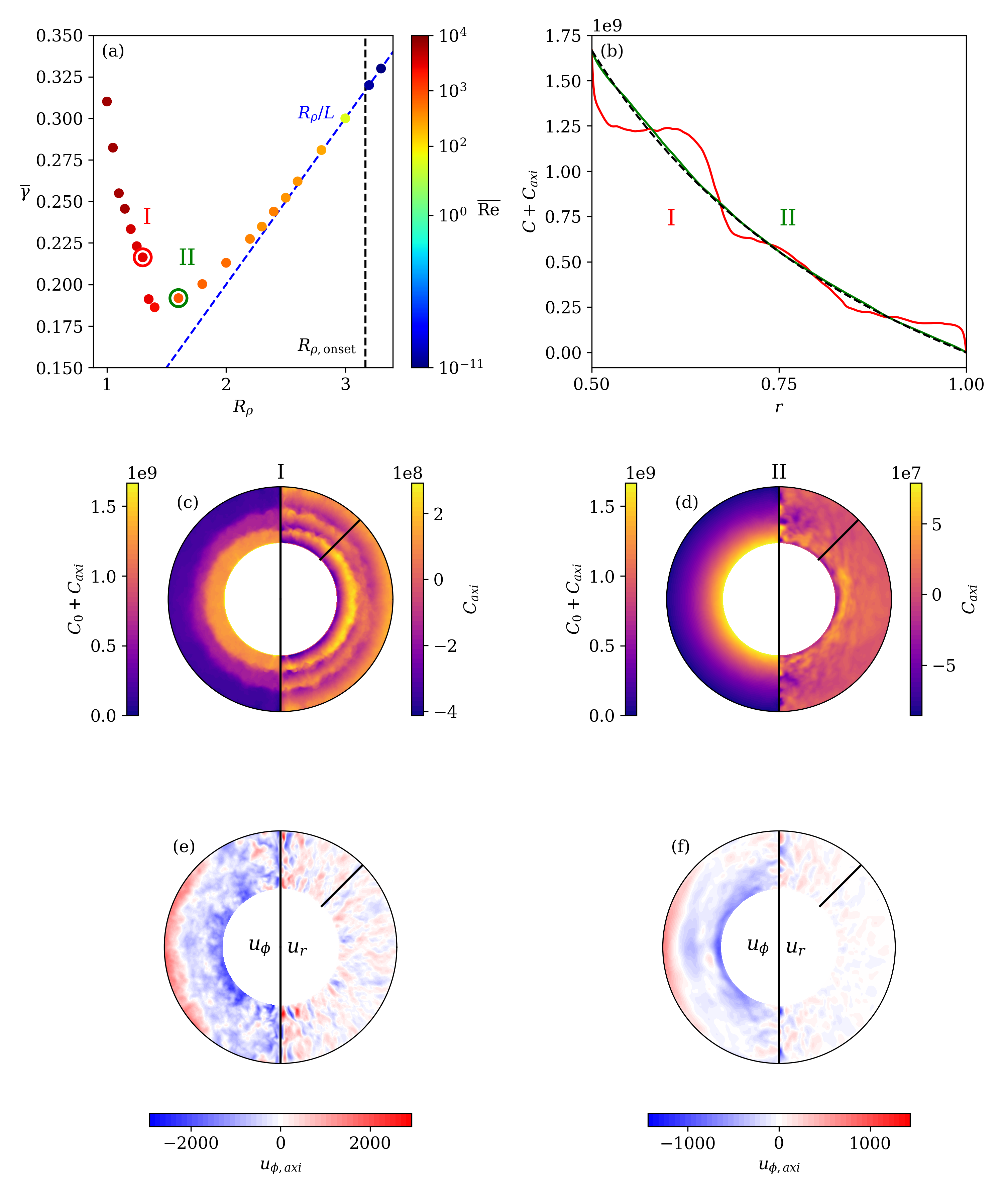}
    \caption{(a) Flux ratio  time-averaged over a short period just after the initial instability has saturated  $\overline{\gamma}$ as a function of $R_\rho$ for simulations along the transect $\Rac=5\times10^9$, $1\leq R_\rho\leq3.3$, with $\Pr=0.3$, $\Sc=3$, $\Delta R=0.5$, $\Ek=10^{-4}$. The values for $\overline{\gamma}$  are calculated (by \eqref{eqn:gamma}) during the growth phase of each simulation. The vertical dashed line marks the limit of rotating semi-convection $R_{\rho,\text{onset}}=3.17$. Points are coloured by the value of $\overline{\R}$ after saturation. The dashed blue line shows the limit of pure conduction $R_\rho/\Le$. (b) Radial profiles of the total axisymmetric ($m=0$) composition field $C_0+C_{axi}$, for the two simulations labelled `I' and `II' in (a). (c),(d) Snapshots of the axisymmetric part of the composition field for the same two simulations, showing both the total field and perturbation; (e),(f) Snapshots of the radial and azimuthal parts of the velocity field. Snapshots are shown immediately after the initial appearance of layers. Black lines in (c)--(f) show the transect along which the profiles in (b) are calculated.}
    \label{fig:gamma}
\end{figure}

As layering is a secondary instability that acts on the initial semi-convection response, $\gamma$ must be calculated immediately after the saturation of the initial instability; measuring the fluxes after the development of layers results in spurious trends in $\gamma(R_\rho)$ due to the different evolution of the layered compared to non-layered simulations. Figure~\ref{fig:gamma}(a) shows how $\overline{\gamma}$ (averaged over a short period just after the saturation of the initial instability) varies as a function of $R_\rho$ across a transect of simulations. 
As $R_\rho$ increases from $1$, $\overline{\gamma}$ decreases, reaching a minimum at $R_\rho\approx1.5$. Past this critical point, $\gamma$ increases almost linearly with $R_\rho$, becoming very close to the conductive limit $\overline{\gamma}\sim R_\rho/\Le$. This curve resembles that found in periodic box simulations by \citet{mirouh2012new} (although with quantitatively different values, likely due to the effect of rotation). Layer formation is predicted for $R_\rho<R_{\rho,\text{layer}}\approx1.5$. The colour of the points denotes the value of $\overline{\R}$, showing that as $R_\rho$ increases, the flow becomes weaker, and a significant drop-off in $\overline{\R}$ occurs past the limit of layering. Note that the two simulations with the largest values of $R_\rho$ are past the onset of rotating semi-convection $R_{\rho,\text{onset}}=\Rac/\RaTc(\Rac)\approx 3.17$ (i.e. are stable to both convection and semi-convection). In these simulations, mass transfer occurs only via conduction, so they lie exactly on the conductive limit.

Figs~\ref{fig:gamma}(b) shows radial profiles of the axisymmetric part of the composition field $C_0+C_{axi}$, for two simulations (one either side of $R_{\rho,\text{layer}}$). Figures~\ref{fig:gamma}(c)--(f) show snapshots of the composition and velocity fields for the same simulations. We show the composition, rather than total density, as fast thermal diffusion (small $\Pr$) means that the temperature profile is rather smeared, making it more difficult to see structures in $\rho$ than in $C$. In simulation I, where $R_\rho<R_{\rho,\text{layer}}\approx1.5$  (fig.~\ref{fig:gamma}(c), (e)), there is a clear layered structure, with three bands of relatively constant composition, separated by sharp interfaces. This structure is also visible in the radial velocity, particularly near the poles, where three different convective bands can be seen.
In simulation II, by contrast, no layers are visible, and the total composition field shows little variation from the background conductive profile $C_0(r)$. Instead, the semi-convection instability results in a small-amplitude linear response that does not develop into layers. The weak velocities seen in fig.~\ref{fig:gamma}(f) result in very small convective fluxes, so the flux ratio is very close to the conductive limit $\overline{\gamma}\sim R_\rho/\Le$.  These two snapshots are representative for all simulations with $R_\rho<1.5$ and $R_\rho>1.5$ respectively, so there is good agreement with the $\gamma$-instability theory, with layers forming where $\partial\overline{\gamma}/\partial R_\rho<0$.

Note that the small departure from the conductive profile for simulation II in fig.~\ref{fig:gamma}(b) does not mean that semi-convection is not occurring; a linear stability analysis (see Appendix \ref{sec:linearonset}) shows that at this value of $\Rac$, the semi-convection limit is $R_{\rho,\text{onset}}=3.17$. For values in the range $R_{\rho,\text{layer}}\lesssim R_\rho<R_{\rho,\text{onset}}$, layers do not develop, and instead the long-time state of the field is a small-amplitude weakly non-linear response, with a Reynolds number (and velocities) two orders of magnitude smaller than in the layered simulation. This matches the results of \citet{moll2016new}, who found that turbulent fluxes remain small for $R_\rho>R_{\rho,layer}$.

\subsubsection{Effect of $\Rat$ on layer formation}\label{sec:layerRac}
As discussed previously, layering in our simulations depends not only on $R_\rho$, but on the magnitude of the thermal forcing $\Rat$ as well, as demonstrated by fig.~\ref{fig:firstsnapshots}. \citet{fuentes2024evolution} suggest that this may be due to geometric constraints, with the instability scale being too large with respect to the domain depth. To investigate the dependence on the stratification, we consider simulations with fixed $R_\rho=1.2$, and vary $\Rac$ and $\Rat$. Thus, the magnitude of the stratification and forcing varies, but the distance from the onset is relatively constant.

\begin{figure}
    \centering
    \includegraphics[width=\textwidth]{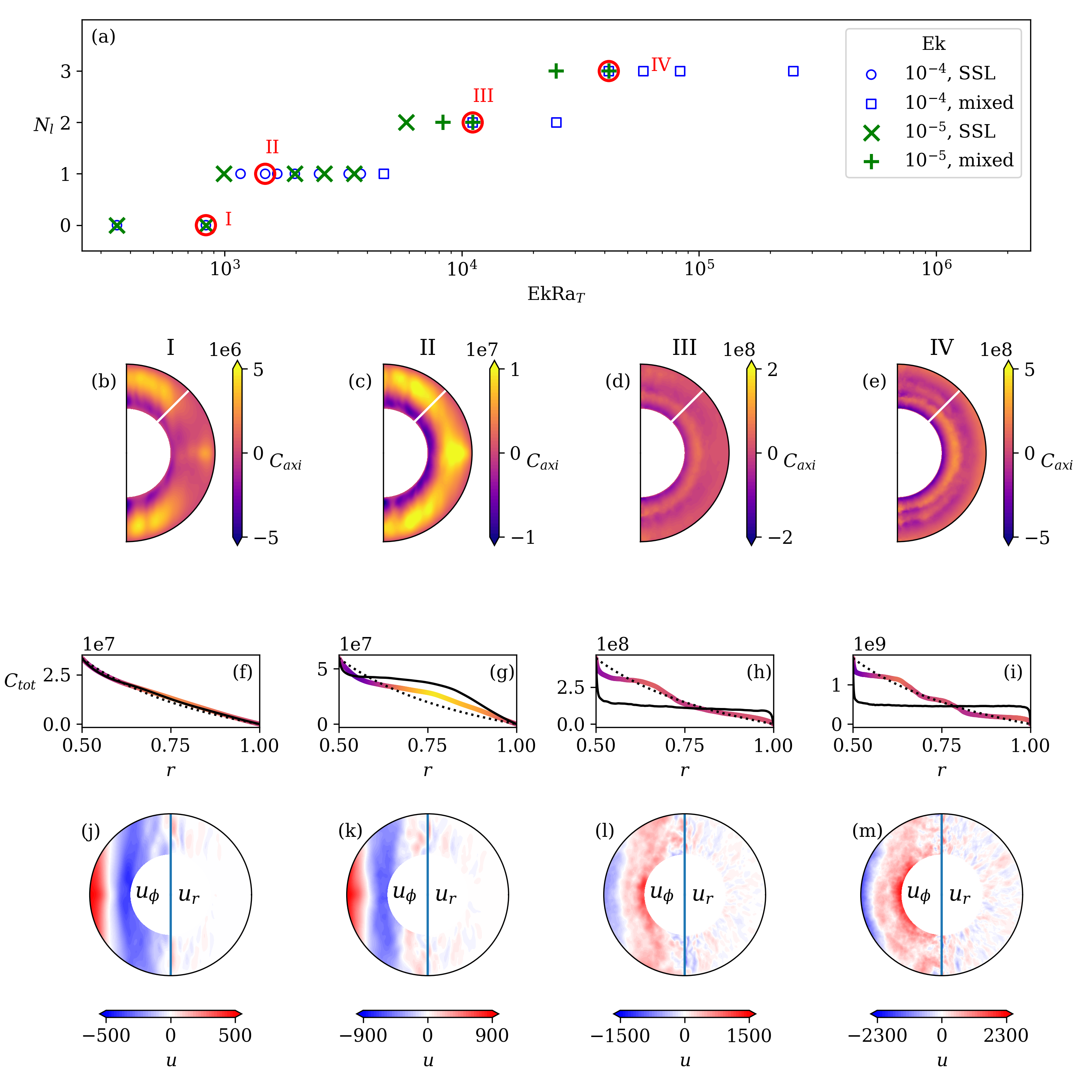}
    \caption{(a) Dependence of the initial number of layers $N_l$ on $\Ek\Rat$, for simulations along transects $R_\rho=1.2$, and $ \Ek=10^{-4}$ and $10^{-5}$, with $\Pr=0.3$, $\Sc=3$, and $\Delta R=0.5$. Circles/crosses represent simulations where the layers eventually decay to a state with a stably stratified layer overlaying a convective zone; squares/plusses represent cases where the final state is a well-mixed region over the majority of the domain. (b)--(m) show snapshots of the simulations from the $\Ek=10^{-4}$ transect labelled ``I''--``IV'' in (a), with each column coming from a single simulation, just after layers have formed. (b)--(e) axisymmetric part of the perturbation composition field $C_{axi}$. (f)--(i) radial profiles of the total axisymmetric composition field $C_{tot}=C_{axi}+C_0(r)$ along the white lines in (b)--(e), coloured by the value of $C_{axi}$. Dotted black lines show the conductive profile $C_0(r)$, while the solid black line shows the final profile of $C_{tot}(r)$ after the end of the layering phase. (j)--(m) axisymmetric part of the azimuthal and radial velocities.}
    \label{fig:Raclayers}
\end{figure}

Figure~\ref{fig:Raclayers} shows how the number of layers $N_l$ depends on $\Rat$ along two transects with $R_\rho=1.2$ for $\Ek=10^{-4}$ and $10^{-5}$, alongside snapshots of simulations with zero, one, two and three layers. The data from the two transects collapses onto a single trend in terms of $\Ek\Rat$ (a parameter that is independent of viscosity). Mixed layers are identified from the composition profiles (fig.~\ref{fig:Raclayers}(f)--(i)) as broad regions where the total composition gradient is notably lower than in the conductive profile $C_0(r)$; where more than one layer exists, they are separated by relatively narrow, high gradient `interfaces'. Figure~\ref{fig:Raclayers}(j)--(m) show the zonal and radial velocities. As discussed in \S~\ref{sec:layerRp}, the layers can also be observed as distinct convective bands in the profiles of $u_r$, although this is significantly less clear than in the composition fields. Over time, the number of layers reduces through a series of mergers; however the timescale for this process is much longer than that for the initial layer formation. As such, the initial number of layers $N_l$ can easily be identified from the snapshots.

To understand this trend in the number of layers, we consider the parameters in terms of the characteristic length $d$  and the modified Taylor number $Ta^*$. The condition of \citet{mirouh2012new} suggests that layering requires $\Rat>(50R_o/\Delta R)^4=10^8$, however for the $\Ek=10^{-4}$ transect we see layering at $\Rat=10^7$, suggesting that the addition of rotation has changed the necessary length scale. According to the $Ta^*\gg1$ condition \eqref{eqn:Ta*}, the critical value of $\Rat$ would scale with $\Ek^2$; however fig.~\ref{fig:Raclayers} shows that the number of layers actually depends on the viscosity-free quantity $\Ek\Rat$, with layering for $\Ek\Rat\gtrsim1000$. We note that our simulations show layering up to $Ta^*=9$, but appear to all be in the small-$Ta^*$ regime; the difference from the $Ta^*=1$ cut-off of \citet{moll2017effect} is likely due to a scaling factor in the definition \eqref{eqn:Ta*}.

Figure~\ref{fig:Raclayers}(a) shows the onset of layering at $\Ek\Rat\approx 1000$, with transitions to two and three layers at $6000$ and $12000$ respectively. Based on this, we identify the empirical scaling $h_{layer}$ for layer formation
	\begin{equation}
		h_{layer}\approx \frac{5}{(\Ek\Rat)^{1/3}},
	\end{equation}
and note that  the number of layers initially present in the domain will be the integer part of $R_o\Delta R/h_{layer}$.

Note the difference between the zero-layer state in fig.~\ref{fig:Raclayers}(b) and that seen in fig.~\ref{fig:gamma}(d). Here, layering is possible by the $\gamma$-instability, so there is a full non-linear response with convective motions visible in the velocity field, and noticeable density structure visible, but no coherent layering throughout the domain. By contrast, in fig.~\ref{fig:gamma}(d), the system is stable to layering, with only a small-amplitude weakly non-linear response. The composition field is rather homogeneous, with no discernible departure from the background conductive state in fig.~\ref{fig:gamma}(d).

\subsection{Long-term evolution of staircases}\label{sec:longtermstaircases}
It is well-documented that in simulations, density staircases gradually erode through layer merger events, until only a single mixed layer remains in the domain \citep[e.g.][]{wood2013new,fuentes2024evolution,pruzina2025onedimensional}. In fig.~\ref{fig:Raclayers}(f)--(i), the solid black lines represent the final profile of $C_{tot}$ after layers have decayed, showing that in our simulations, there are two possible final states: low values of $\Rat$ lead to a structure comprised of an inner convective region overlain by a wide stably stratified layer, while higher values of $\Rat$ produce a single well-mixed region filling the domain, with narrow boundary layers on either side. 

\begin{figure}
    \centering
    \includegraphics[width=\textwidth]{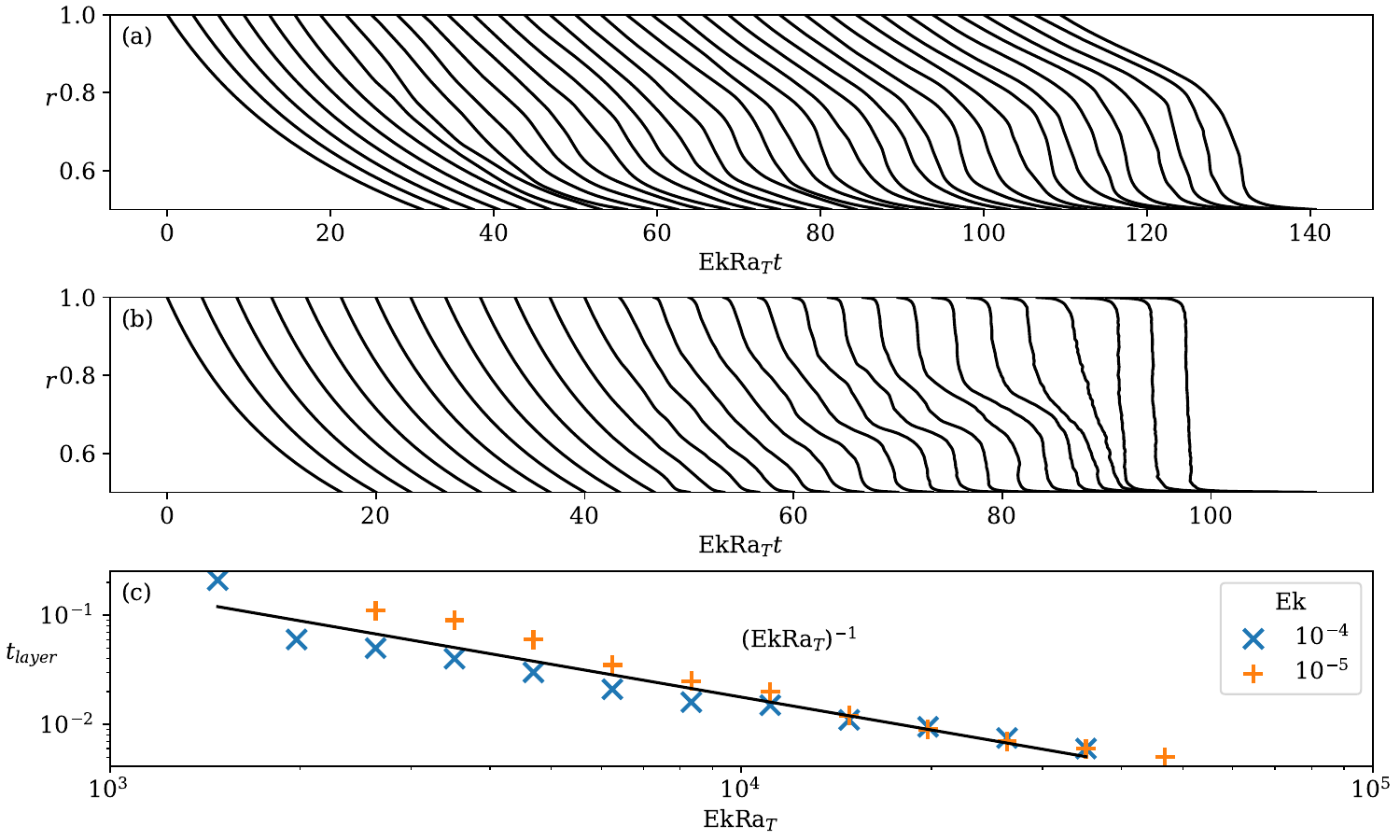}
    \caption{Profiles of $C_{tot}(r)$ throughout the layered phase of two simulations, for (a) $\Rat = 1.15\times10^7$, (b) $\Rat = 4.2\times 10^8$. Profiles are shown at a series of times, which can be read at the top left of each profile. The horizontal ($C$) scale for each profile is the same due to the fixed-$C$ boundary conditions. (c) Time spent in layered state $t_\text{layer}$, for simulations in the final convective zone regime.}
    \label{fig:layerevolution}
\end{figure}

Figure~\ref{fig:layerevolution} shows the long-term evolution of the radial composition profile for two simulations: one leading to each final state. In fig.~\ref{fig:layerevolution}(a), for the (low-$\Rat$ simulation labelled  ``II'' in fig.~\ref{fig:Raclayers}), the initially smooth background profile gradually sharpens, leading to a `one-layer' state, with a single well-mixed layer in the interior, overlain by a stably stratified region. Over time, the mixed layer increases in size, but eventually stabilises far from the outer boundary. This state is remarkably stable for long times. By contrast, in fig.~\ref{fig:layerevolution}(b), (simulation ``IV'' in fig.~\ref{fig:Raclayers}), the initial profile develops into a stack of three mixed layers separated by sharper interfaces, which merge until a single mixed layer takes up the vast majority of the domain, between narrow stably stratified boundary layers. 

For higher values of $\Rat$, the behaviour of the layers is similar to that seen in other studies, with a gradual erosion of the layered state leading ultimately to a well-mixed state throughout the domain \citep[e.g.][]{fuentes20253d}. However, the final state seen for lower values of $\Rat$ is rather different. This suggests a competition between rotational and convective processes, where the tendency of a rotating fluid to form columnar structures prevents the erosion of the stably stratified outer layer. For these low-$\Rat$ simulations, the layers never fully merge --- rather, the wide convective layer overlain by an SSL survives without further evolution of its size --- meaning that it can impact the dynamics on long time scales. 

Figure~\ref{fig:layerevolution}(c) shows the survival time of the layered state, for simulations resulting in a convective zone throughout the entire domain (it is difficult to define an end of the layered phase when the final state contains a wide SSL). Larger values of $\Ek\Rat$ result in a shorter layered phase, with an approximate scaling $t_\text{layer}\sim(\Ek\Rat)^{-1}$, independent of the viscosity. This agrees with the results of \citet{fuentes2024evolution}, who found that higher rotation rates relative to the stratification produced longer-lasting layers, as rotational effects resisted the convection processes that cause layers to merge.

To quantify the transition between the two regimes, we define the boundaries of the convective zone as 
\begin{eqnarray}
    R_1 &=& \min \{r | N^2(r)< 0\},\label{eqn:R1}\\
    R_2 &=& \max \{r | N^2(r)< 0\}.\label{eqn:R2}
\end{eqnarray}
We identify the height of the inner and outer stably stratified layers as
\begin{equation}
    h_{in}=R_1-R_i,\quad h_{SSL}=R_o-R_2.
\end{equation}

\begin{figure}
    \centering
    \includegraphics[width=\textwidth]{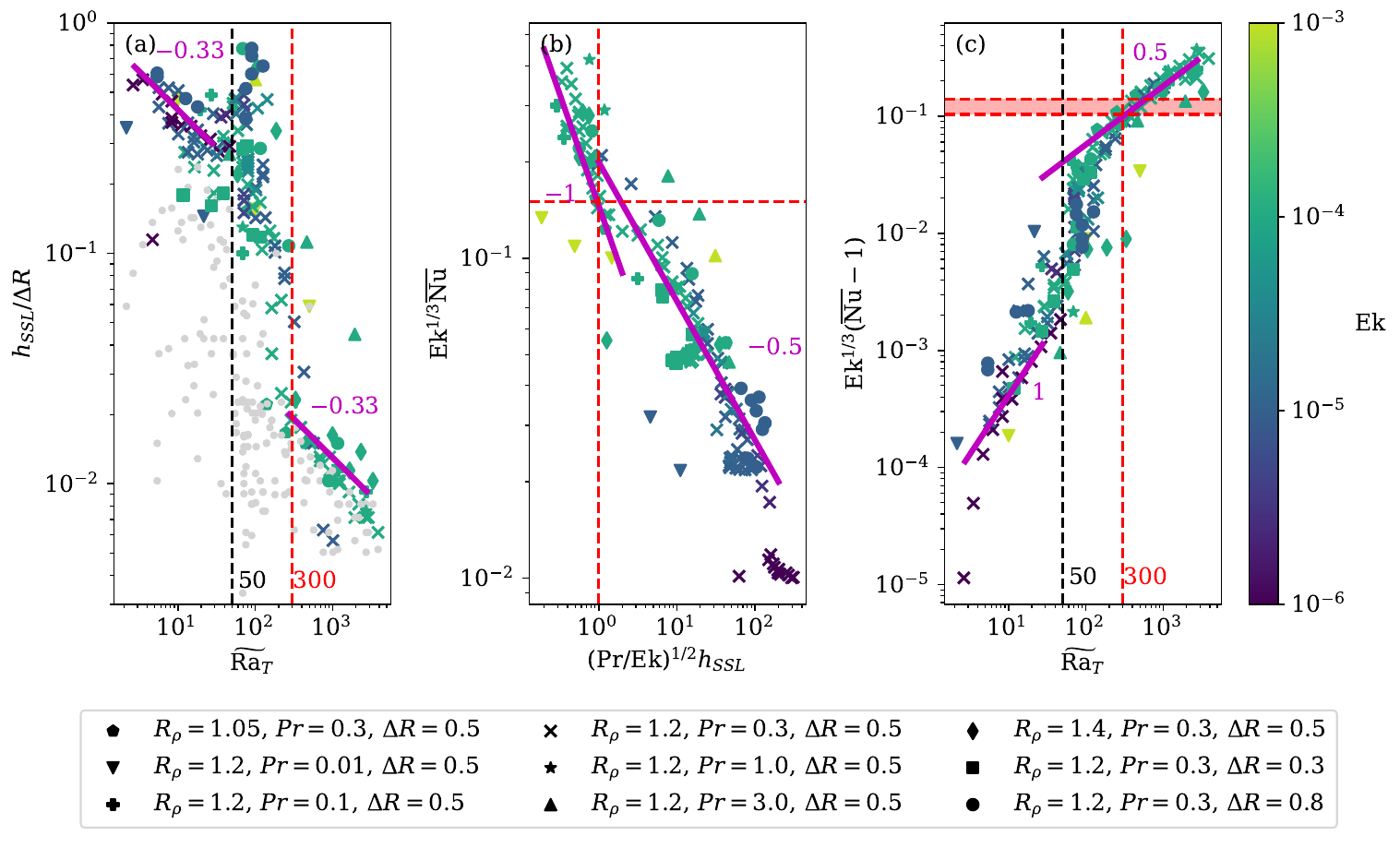}
    \caption{ (a) Dependence of the height $h_{SSL}$ of the outer stably stratified layer on $\Rred$; (b) variation of the Nusselt number at the outer boundary as a function of the rescaled SSL height $(\Pr/\Ek)^{1/2}h_{SSL}$; (c) dependence of the time-averaged Nusselt number $\overline{\Nu}$ on $\Rred$. Points are coloured by Ekman number $\Ek$, with different values of $R_\rho$, $\Pr$ and $\Delta R$ represented by different symbols. Grey points in panel (a) show the width $h_{in}$ of the inner boundary layer.  The black dashed line in (a) and (c) marks a change in scaling for $h$, corresponding with a jump in $\overline{\Nu}$. The red dashed lines in all panels correspond to a transition as $(Pr/\Ek)^{1/2}h_{SSL}$ passes through unity: in (b), the horizontal line shows the value of $\Ek^{1/3}\overline{\Nu}$ at this point; in (c), the red shaded band shows this value of $\overline{\Nu}$ for the range of $\Ek$ surveyed, and the vertical line shows the value  $\Rred\approx300$ at which these lines cross the corresponding transects. In panel (a), the vertical red line shows that this value of $\Rred=300$ corresponds to a change of scaling for $h_{SSL}(\Rred)$.}
    \label{fig:innerlayerwidth}
\end{figure}

Figure~\ref{fig:innerlayerwidth}(a) shows how $h_{SSL}$ varies as a function of the thermal forcing, for a range of simulations with $10^{-6}\leq\Ek\leq10^{-4}$. The majority of these simulations take $R_\rho=1.2$, $\Pr=0.3$, and $\Delta R=0.5$, but these values were varied in a small number of simulations, while keeping $\Le=10$ constant. The data  collapse relatively well onto a single curve with $h_{SSL}$ controlled by the  `reduced Rayleigh number' 
\begin{equation}
    \Rred = \Ek^{4/3}\Rat,
\end{equation}
in which $\Rat$ is rescaled by the asymptotic scaling for the onset of rotating convection \citep{dormy2004onset}. This parameter measures the distance from the rotating convection onset, with smaller values of $\Rred$ being closer to the onset, and therefore more strongly rotationally constrained. 
It is also possible to fit the data to other rescaled parameters; for example, there is a similarly good collapse in terms of the control parameter $\Ek^{3/2}\Rat\Pr^{-1}$, which does not depend on length. We note that a similar parameter $\Ek^{3/2}\Rat$ is suggested by \citet{king2012heat} to control the transition between rotating and non-rotating convection, for $\Pr\geq1$.

The grey points in fig.\ref{fig:innerlayerwidth}(a) show the value of $h_{in}$, which generally takes values significantly smaller than $h_{SSL}$, but follows a similar trend. Three regimes are clearly visible. For $\Rred<50$, $h_{SSL}/\Delta R$ is rather large, taking a value between $0.2$ -- $0.7$, and scales rather weakly, with $h_{SSL}/\Delta R\sim\Rred^{-1/3}$.  Above $\Rred\gtrsim300$, the same $-1/3$ power law is seen. Between these regimes, there is a transition region where $h_{SSL}$ varies quickly as a function of $\Rred$, with significant spread in the data meaning that dedicated study would be necessary to further classify the trend.

These transitions can be explained in terms of the dependence of $\Nu$ on $h_{SSL}$. In Rayleigh B\'enard convection, the thermal boundary layer thickness is conventionally defined as $\delta_T=H/(2\Nu)$, where $H$ is the depth of the domain \citep[see][]{belmonte1994temperature}. Figure~\ref{fig:innerlayerwidth}(b) shows that, in our simulations, $\overline{\Nu}\sim h_{SSL}$ for $h_{SSL}\lesssim (\Ek/\Pr)^{1/2}$. This cut-off corresponds to the length above which the Coriolis force acts faster than thermal diffusion. At this point, $\Ek^{1/3}\overline{\Nu}\approx0.15$, which is shown as a red shaded band on fig.~\ref{fig:innerlayerwidth}(c) for the range of $\Ek$ surveyed. The data falls in this band at $\Rred\approx300$; the same value at which the kink is visible in fig.~\ref{fig:innerlayerwidth}(a).
At these small scales, rotation has only a small effect compared with diffusion, so a scaling independent of $\Ek$ may be expected. In fig.~\ref{fig:innerlayerwidth}(a), the shift between the $\Ek=10^{-4}$ and $10^{-5}$ transects for $\Rred>300$ is consistent with the rotation-free scaling $h_{SSL}\sim\Rat^{-1/3}$, matching that expected for a boundary layer in thermal convection \citep[e.g.][]{king2012heat}.

When the SSL is thicker, fig.~\ref{fig:innerlayerwidth}(b) shows a shallower slope $\overline{\Nu}\sim h_{SSL}^{-1/2}$. The transition at $\Rred\approx50$ does not have a clear signature in \ref{fig:innerlayerwidth}(b), with the $\overline{\Nu}(h_{SSL})$ dependence remaining the same. On the other hand, fig.~\ref{fig:innerlayerwidth}(c) shows a sharp jump of more than an order of magnitude in the value of $\overline{\Nu}-1$ at $\Rred=50$, which can be accounted for as a transition to turbulence, occurring at the same point as the first transition in fig.~\ref{fig:innerlayerwidth}(a).

We note that a boundary layer interpretation is also possible for $\Rred<50$, with the scaling $h_{SSL}\sim\Rred^{-1/3}$ matching that obtained for a boundary layer in rotating thermal convection \citep{king2012heat}. This suggests the SSL obtains the maximum size of a thermally unstable layer before the onset of convection. After the initial semi-convection instability the system stabilises into the largest possible non-convecting region, with a convective zone beneath.

To summarise, the SSL generally scales as a thermal boundary layer. When $\Rred<50$, the flow is non-turbulent, $h_{SSL}$ follows the scaling $\Rred^{-1/3}$ for rotating thermal convection. For $\Rred>300$, the thermal forcing outweighs rotation, and the SSL takes a scaling for non-rotation convection $\Rat^{-1/3}$. In between these limits, there is a transition range where the flow is both turbulent and influenced by rotation, and $h_{SSL}$ quickly transitions between the non-rotating and rotating boundary layers.  

\subsection{Behaviour of the saturated semi-convection field }
We now consider the dynamics beyond the layered phase, when the system has saturated to a statistically steady state.  In \S~\ref{sec:longtermstaircases} we noted two regimes, with the post-layered state consisting of either a single convective region, or a convective zone overlain by a stably stratified layer with a mostly zonal flow. To understand the signature of these regimes on the global properties of the flow, we define Reynolds numbers based on the 
zonal-toroidal ($u_\text{ZT}$), and non-zonal-poloidal ($u_\text{NZP}$) components of the velocity (from the poloidal-toroidal decomposition used by XSHELLS). The radial velocities are poloidal only and the zonal jets are toroidal only. Hence, the zonal-toroidal and non-zonal-poloidal components can be used as proxies for the strength of zonal jets and convective motions, respectively:
\begin{equation}
    \R_\text{jet} = \frac{\langle u_\text{ZT}\rangle_\text{rms}R_o}{\nu},\quad \R_\text{conv} = \frac{\langle u_\text{NZP}\rangle_\text{rms}R_o}{\nu}.
\end{equation}

\begin{figure}
    \includegraphics[width=\textwidth]{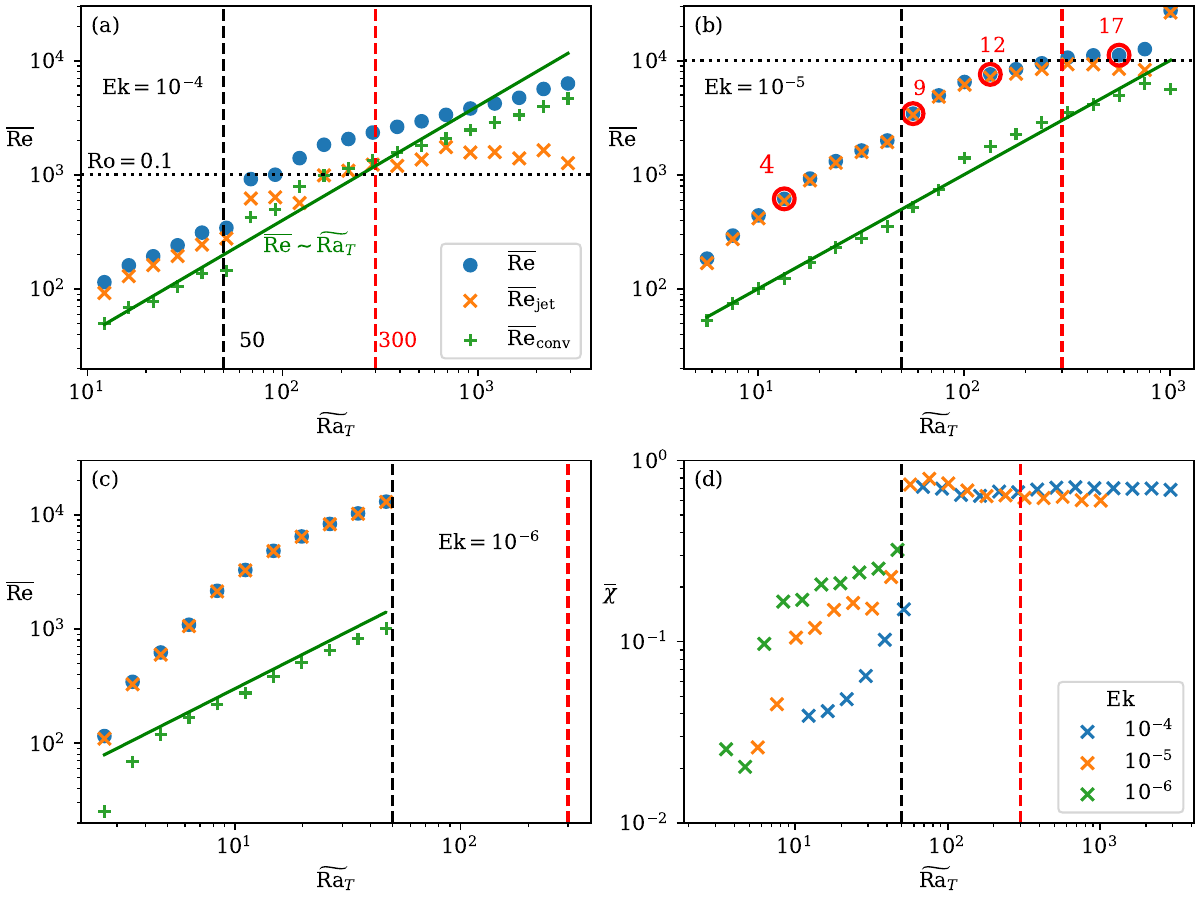}
    \caption{Time-averaged Reynolds number $\overline{\R}$ plotted as a function of $\Rred$ for simulations with $R_\rho=1.2$, $\Pr=0.3$, $\Sc=3$, $\Delta R = 0.5$, for (a) $\Ek=10^{-4}$, (b) $10^{-5}$ and (c) $10^{-6}$. Blue points show the total Reynolds number, with orange and green showing the jet- and convection-based Reynolds numbers $\overline{\R}_\text{jet}$ and $\overline{\R}_\text{conv}$ respectively. Red circles mark simulations that are shown in detail in fig.~\ref{fig:transition}. Red and black dashed lines mark the positions of the transitions seen in fig.~\ref{fig:innerlayerwidth}, and green lines show the scaling $\R_\text{conv}\sim\Rred$. Black dashed line shows $\Ro=0.1$ (d) Mixing efficiency $\overline{\chi}$ calculated in the long-term saturated state, plotted for three values of $\Ek$.}
        \label{fig:ReRac}
\end{figure}

Figure~\ref{fig:ReRac}(a)--(c) shows the dependence of $\overline{\R}$, $\overline{\R}_\text{jet}$ and $\overline{\R}_\text{conv}$ on $\Rred$ along the transect $R_\rho=1.2$, for $\Ek=10^{-4}$, $10^{-5}$ and $10^{-6}$. In general, as the forcing increases, the flow strength also increases. In \S~\ref{sec:longtermstaircases} we noted two visible transitions. These transitions are also visible in fig.~\ref{fig:ReRac}, with a sudden jump in the value of $\overline{\R}$ at $\Rred = 50$, and a saturation of the value of $\overline{\R}_\text{jet}$ at $\Rred\approx300$. For the majority of simulations, $\overline{\R}_\text{jet}\gg\overline{\R}_\text{conv}$, with convective motions being more dominant only for the $\Ek=10^{-4}$ transect where $\Rred\gtrsim
300$. Here, the outer stably stratified layer is thinner than $(\Ek/\Pr)^{1/2}$ (see fig.~\ref{fig:innerlayerwidth}). At fixed $\Ek$, there is a relatively good fit to the scaling $\R_\text{conv}\sim\Rred$, particularly for the smaller values of $\Ek$.
Hereafter, we will refer to the regime with $\Rred<300$ as ``jet-dominated", and $\Rred>300$ as ``convection-dominated".

Figure~\ref{fig:ReRac}(d) shows the mixing efficiency in the convective zone, calculated \cite[following][]{vidal2018magnetic} as 
\begin{equation}
    \chi=1-\frac{\int_{R_1}^{R_2}\langle N^2\rangle_{S(r)} \mathrm{d}r}{\int_{R_1}^{R_2}\langle N^2_0\rangle_{S(r)} \mathrm{d}r},
\end{equation}
where $\langle\cdot\rangle_{S(r)}$ is the mean over the spherical surface at radius $r$, and $R_1$ and $R_2$ are defined by \eqref{eqn:R1}--\eqref{eqn:R2}. This quantifies how well-mixed the convective zone is, with $\chi=0$ when there is no departure from the background density profile, and $\chi=1$ when the convective zone is fully mixed to a constant density. Fig.~\ref{fig:ReRac}(d) shows that, in the jet-dominated regime, $\overline{\chi}$ takes small values, but increases as a function of $\Rred$. By contrast, $\overline{\chi}$ is large and relatively constant in the convective regime. This means that not only is the convective zone larger for values of $\Rred>50$, but the convection itself is also more efficient for mixing. 

To quantify the effect of rotation on the flow, it is common to define the Rossby number
\begin{equation}
	\Ro=\Ek\R.
\end{equation}
When $\Ro\lesssim0.1$, flows are considered to be rotationally constrained; above this limit, rotational effects are much less important. In figs.~\ref{fig:ReRac}(a) and (b), $\Ro=0.1$ is marked with horizontal dotted lines; we see that for both values of $\Ek$, as the forcing increases, the $\R_\text{jet}$ remains close to $\R$ until this cut-off is reached, but for $\Ro>0.1$, the jet-based Reynolds number reaches a plateau. This cut-off corresponds relatively well with the jet-convection transition already identified at $\Rred\approx300$. We note that the other transition, at $\Rred\approx50$, occurs for $\Ro=0.03$. For the smallest value of $\Ek$ surveyed, all simulations are strongly rotationally constrained, with a maximum value of $\Ro\approx 0.01$.

Figure \ref{fig:ReRac} does not demonstrate a unified behaviour across different values of $\Ek$ and the whole range of $\Rred$; we shall discuss how data collapses on a single master curve later in \S~\ref{sec:dynamo}.

\begin{figure}
    \includegraphics[width=\textwidth]{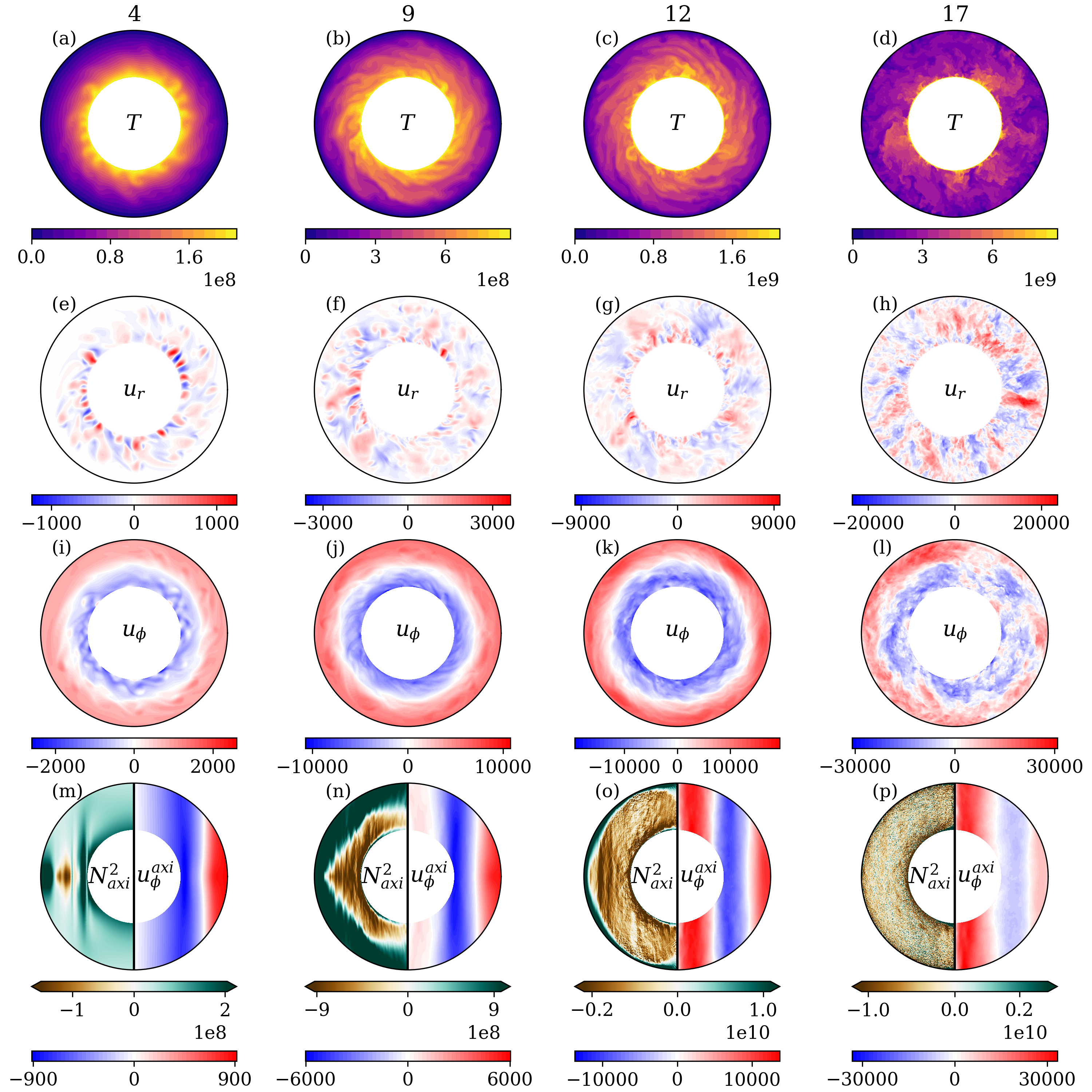}
    \caption{Snapshots of the four simulations marked by red circles in fig.~\ref{fig:ReRac}(b). Each column shows results from a single simulation (labelled at the top). Each row shows, respectively, equatorial slices of the temperature perturbation $T$, radial velocity $u_r$, azimuthal velocity $u_\phi$, and meridional slices of the axisymmetric ($m=0$) part of the azimuthal velocity $u_\phi^{axi}$ and axisymmetric squared Brunt-V\"ais\"al\"a frequency $N^2_{axi}$.}
    \label{fig:transition}
\end{figure}

The transitions at $\Rred=50$ and $300$ are most obvious in the $\Ek=10^{-4}$ transect, although are also present in the simulations with $\Ek=10^{-5}$.  To look deeper into the nature of these transitions, fig.~\ref{fig:transition} shows snapshots of the velocity and temperature fields, for four simulations labelled by their position along the $\Ek=10^{-5}$ transect (cf. fig.~\ref{fig:ReRac}(b)). 

For $\Rred<50$ (labelled `4' in fig.~\ref{fig:ReRac}(b)), the flow is generally weak and laminar, with a prograde equatorial jet, retrograde flow at higher latitudes, and some wave-like structures along the shear layer. The flow here is strongly rotationally constrained, with $\Ro=0.006$. The density perturbation is also rather weak, with a stable stratification retained throughout almost the entire domain, except a small patch near the equator. For simulation 9, with $\Rred>50$, the flow is turbulent, with figs.~\ref{fig:transition}(b),(f), and (j) showing much more intense flows and greater disruption of the temperature field than in simulation 4. The effect of rotation is still very strong, with $\Ro=0.03$. Figure~\ref{fig:transition}(n) shows that for simulation 9, the zonal flow has a three-jet structure, with prograde flow at the poles and near the equator, and retrograde flow in between. The density field is separated into two clear layers; an inner convective zone with $N^2_{axi}<0$, and an outer SSL with $N^2_{axi}>0$.

For simulation 12, for a higher value of $\Rred$, but still beneath the jet-convection transition, the only major difference from simulation 9 is that the outer stably stratified layer is significantly narrower.

Finally, for $\Rred>300$, simulation 17 shows very strong convective flows.  Figure~\ref{fig:transition}(l) shows that the zonal jets are rather broken up compared to the other simulations, and fig.~\ref{fig:transition}(p) shows that the zonal flow is on average weaker and the shear layers more smeared. Here, the effect of rotation is much weaker than the other simulations, with $\Ro=0.11$. The profile of $N^2_{axi}$ shows that the convective zone fills the entire domain, with no SSL visible anywhere.
Note that while the profile of $N^2$ appears rather noisy, the colourbar has been chosen to highlight changes of sign (to clearly show the border of the convective and stable layers). In a well-mixed state, $N^2$ is dominated by small-scale fluctuations around zero, resulting in the noisy profile visible in fig.~\ref{fig:transition}(p).

\begin{figure}
    \centering
    \includegraphics[width=\textwidth]{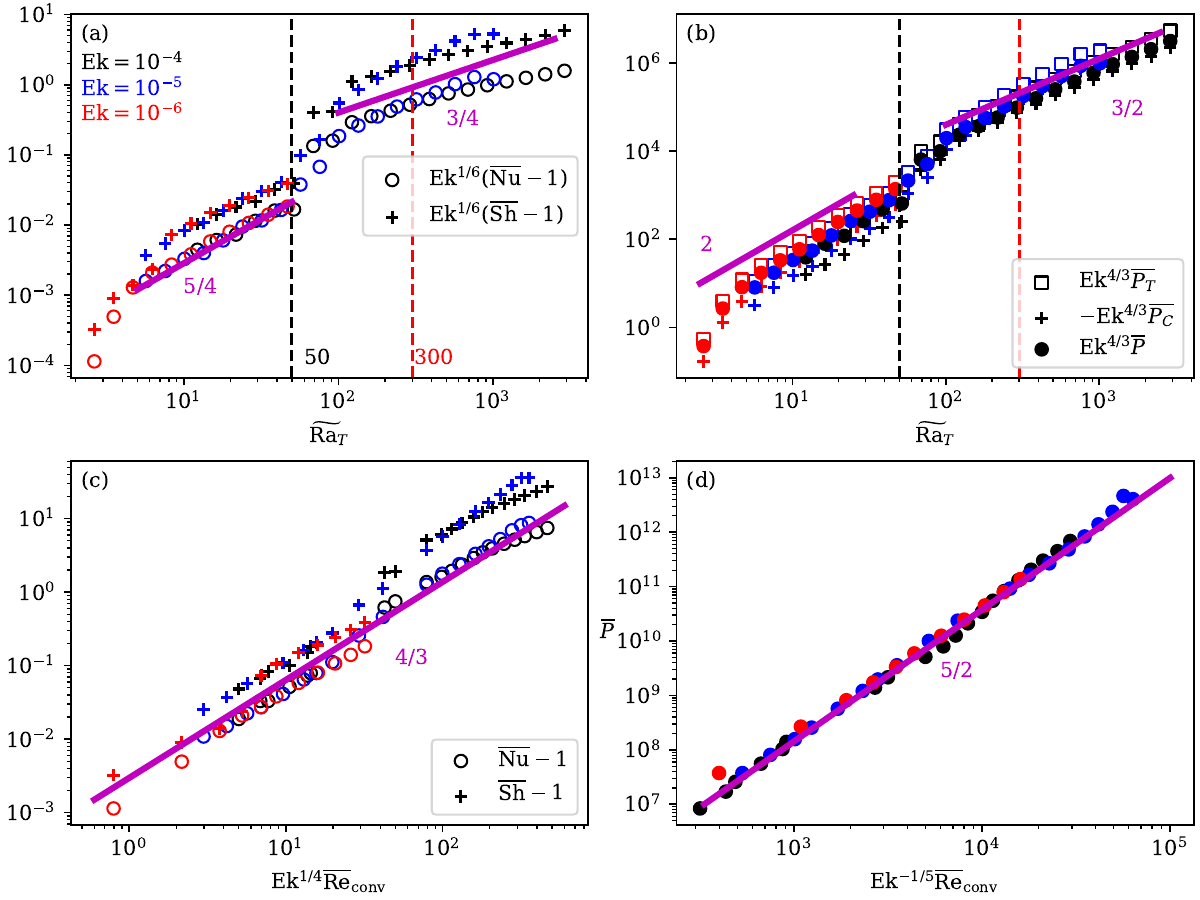}
    \caption{Quantities describing the strength of convection for simulations along the transect $R_\rho=1.2$: (a) Time-averaged Nusselt and Sherwood numbers on the outer boundary $r=R_o$, as a function of $\Rred$. (b) Time-averaged thermal, compositional and total convective power. Vertical dashed lines denote the positions of the transitions identified in fig.~\ref{fig:innerlayerwidth}. (c) Nusselt and Sherwood numbers plotted against convective Reynolds number $\R_\text{conv}$, scaled by $\Ek^{1/3}$. (d) Convective power plotted against $\Ek^{-1/3}\R_\text{conv}$.}
    \label{fig:NuCP}
\end{figure}

To quantify the turbulent transport in the flow, we  show the trend in the Nusselt and Sherwood numbers on the outer boundary, as a function of $\Rred$. Figure~\ref{fig:NuCP}(a) shows that, at all points, $\overline{\Sh}-1$ is larger than $\overline{\Nu}-1$ (as the diffusive transport is naturally lower for the slower-diffusing scalar). For smaller values of $\Rred$, $\overline{\Nu}$ and $\overline{\Sh}$ are both close to unity, representing very low levels of convection. As $\Rred$ increases, $\overline{\Nu}$ and $\overline{\Sh}$ increase, scaling as $(\overline{\Nu}-1),(\overline{\Sh}-1)\sim\Ek^{-1/6}\Rred^{5/4}=\Ek^{3/2}\Rat^{5/4}$. 
The influence of rotation here is relatively small, and the power $-1/6$ is not exact; a relatively good fit is found for $\Ek^{-0.16\pm0.1}$. For comparison, \citet{wood2013new} found $\overline{\Nu}-1\sim\Rat^{1/3}$  for non-rotating, layered semi-convection. For rapidly rotating thermal convection,  \citet{gastine2016scaling} found the upper limit on the power law to be $\overline{\Nu}\sim \Ek^2\Rat^{3/2}/\Pr^{1/2}$, while near to the onset of rotating convection $\Rat^c$, the scaling was $\overline{\Nu}-1\sim\Rat/\Rat^c-1$. The $\Rat^{5/4}$ law that we obtain sits between these two limits. 

The transition at $\Rred=50$ is clearly visible in fig.~\ref{fig:NuCP}(a), with a marked jump of a factor of ten in both $\overline{\Nu}-1$ and $\overline{\Sh}-1$, and a change in the scaling law to $\Ek^{1/6}(\overline{Nu}-1)\sim\Rred^{3/4\pm1/4}$. In the fingering regime, \citet{monville2019rotating} found that $\overline{\Nu}-1$ always remained small, meaning that the thermal transport was dominated by diffusion, allowing small-scale fingering patterns to develop. Here, by contrast, $\overline{\Nu}-1$ achieves values larger than unity  due to the vigorously convecting layer for larger values of $\Rred$.

The trend in the convective power $P=P_T+P_C$ against $\Rred$, as well as the components $P_T$ and $P_C$, is shown in fig.~\ref{fig:NuCP}(b). The data collapses to a master curve $\Ek^{4/3}P\sim\Rred^2$ at low values of $\Rred$, with the slope shallowing to $\Rred^{3/2}$ above the transition to turbulence at $\Rred=50$. Upon closer inspection, it can also be seen that, for $\Rred<50$, the total power $\overline{P}=\overline{P_T}+\overline{P_C}$ is very close to $\overline{P_T}$, with a small contribution from $\overline{P_C}$. However, for $\Rred>50$, $-\overline{P_C}$ becomes closer to $\overline{P_T}$, leading to a relative reduction in $\overline{P}$ compared to $\overline{P_T}$.
This matches the description of a convection-dominated flow, where both thermal and compositional fields are strongly disrupted by fluid motions, giving a value of $\overline{P_C}$ that is close to $-\overline{P_T}$.  Notably, the transition at $\Rred=300$ is not visible in either the Nusselt/Sherwood numbers or the convective powers, similarly to what was observed in the flow snapshots.

Figures~\ref{fig:NuCP}(c) and (d) show, respectively, the dependence of the Nusselt and Sherwood numbers, and the convective power, on the convective Reynolds number $\R_\text{conv}$. Fairly simple scalings are visible, with $\overline{\Nu}-1$ and $ \overline{\Sh}-1\sim\left(\Ek^{1/4}\overline{\R}_\text{conv}\right)^{4/3}$, and $\overline{P}\sim\left(\Ek^{-1/5}\overline{\R}_\text{conv}\right)^{5/2}$. These simple scalings, based on $\R_\text{conv}$, rather than $\R$, and in particular showing no significant change at the jet-convection transition, confirm that the majority of turbulent transport is done by convection, with a smaller contribution from the zonal jets.

\section{Dynamo generation by semi-convection}\label{sec:dynamo}

To investigate the possibility of dynamos driven by semi-convection, we now consider the full magneto-hydrodynamic Boussinesq equations, allowing $\bm{b}$ to be non-zero in  \eqref{eqn:momentumequation}--\eqref{eqn:bequation}. We begin by discussing the favourable conditions for the dynamo to exist. The relative strength of the advection to diffusion terms in \eqref{eqn:bequation} can be characterised by the magnetic Reynolds number
\begin{equation}
    \Rm = \frac{\langle u\rangle_\text{rms}R_o}{\eta} = \Pm\R,
\end{equation}
where $\Pm$ is the magnetic Prandtl number (cf. table~\ref{tab:parameters}). When $\Rm\gg1$, advection dominates magnetic diffusion, and dynamo action is  possible. In planetary conditions $\Pm\ll1$, which imposes that $\R$ must be large to produce a self-sustaining dynamo. Referring to fig.~\ref{fig:ReRac}, this means that more intense forcing (i.e. larger values of $\Rat$) is preferable. 

Dynamos are frequently classified in terms of the ratio of the magnetic energy to the kinetic energy, defined in dimensionless form as 
\begin{equation}
	E_b=\frac{1}{2}\bm{b}\cdot\bm{b},\quad	E_u = \frac{1}{2}\bm{u}\cdot\bm{u}.
\end{equation}
When $E_b/E_u>1$, the dynamo is said to be in the strong field regime; in this case, the magnetic field has an important feedback into the velocity. For $E_b/E_u<1$, in the weak field regime, the magnetic field has a weaker effect on the flow.

\citet{pruzina2025planetary} demonstrated that semi-convection can indeed lead to dynamo action at low magnetic Prandtl number. However, this initial study left many outstanding open questions. Notably, this paper focused on a single simulation, producing results relevant to a Jupiter-like configuration. A broader study is still required to understand how the form of the magnetic field depends on the flow structure and dimensionless parameters.

The simulation discussed by \citet{pruzina2025planetary} was in the jet-dominated regime, with the final state of the layered phase being an inner convective zone overlain by a stably stratified region. Fluid motions in this convective zone drive the dynamo to produce a magnetic field. Velocities are lower in the outer layer, although this alone does not necessarily prevent that region from also actively contributing to the dynamo \citep{avalos2003}.

To investigate the onset of the dynamo, we begin from the hydrodynamic simulations presented in \S~\ref{sec:non-lineardynamics}. Taking the saturated semi-convective state as an initial condition, we seed a random small-amplitude magnetic field perturbation and allow the system to evolve in time. After a short adjustment period, the magnetic field undergoes a exponential growth/decay phase; we identify a successful dynamo based on sustained growth of $E_b$.

\subsection{Dynamo action: onset and self-sustained regime}
\begin{figure}
    \centering
    \includegraphics[width=\textwidth]{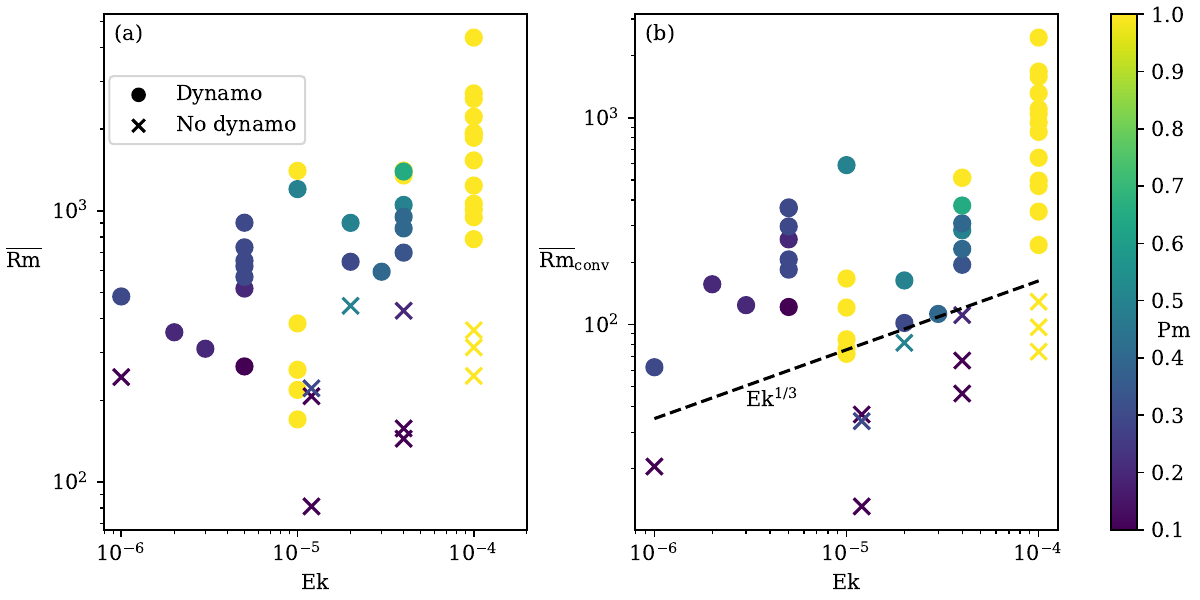}
    \caption{Summary of a large number of dynamo simulations, for $\Pr=0.3$, $\Sc=3$, $R_\rho=1.2$, and a range of values of $\Ek$, $\Rat$, $\Rac$ and $\Pm$, with (a) $\overline{\Rm}$ and (b) $\overline{\Rm}_\text{conv}$ plotted against $\Ek$. Dots and crosses (coloured by $\Pm$) designate simulations with successful and unsuccessful dynamos, respectively. Black dashed line in (b) represents an approximate scaling for the onset of dynamo.}
    \label{fig:dynamoonset}
\end{figure}

Figure~\ref{fig:dynamoonset}(a) summarises a number of simulations for a range of $\Ek$, $\Rat$, $\Rac$ and $\Pm$, showing the value of $\Rm$ achieved, and whether or not a dynamo developed. It is clear that in general, larger values of $\overline{\Rm}$ produce dynamo action; although the onset varies strongly with $\Ek$ and $\Pm$. However, defining the magnetic Reynolds number based on the full domain neglects the fact that the convective flow is rather weak in the outer stably stratified layer. We instead define a convective magnetic Reynolds number. 
\begin{equation}
    \Rm_\text{conv}= \Pm\R_\text{conv}\left(\frac{R_o^3-R_i^3}{R_2^3-R_1^3}\right)^{1/2}\frac{R_2-R_1}{R_o-R_i}.
\end{equation}
where $R_1$ and $R_2$ are the inner and outer boundaries of the convective zone as defined by \eqref{eqn:R1}--\eqref{eqn:R2}, and $R_i$ and $R_o$ are the inner and outer boundaries of the domain (such that $h_{SSL}=R_o-R_2$). This new parameter is based on an assumption that the convective part of the kinetic energy is contained entirely in the convective region, introducing a factor of $\left((R_o^3-R_i^3)/(R_2^3-R_1^3)\right)^{1/2}$ into the r.m.s. velocity. The factor of $(R_2-R_1)/(R_o-R_i)$ is due to rescaling the length in the definition of $\R$.

Figure~\ref{fig:dynamoonset}(b) shows the same data as panel (a), plotted with this new convective magnetic Reynolds number $\overline{\Rm}_\text{conv}$. Here, there is a more consistent trend in the onset of dynamo action across the values of $\Ek$, confirming that the dynamo is only active in the convective zone, and not in the SSL. For the range of $\Ek$ studied, the critical value of $\Rm_\text{conv}$ for dynamo action scales approximately with $\Ek^{1/3}$. This is a favourable regime for a geophysical setting, suggesting only small $\Rm$ may be necessary for dynamo when planetary values of $\Ek$ are considered. However, we note that the trend cannot hold to very low values of $\Ek$, otherwise the critical value of $\Rm$ could become arbitrarily small.

\begin{figure}
\centering
    \includegraphics[width=\textwidth]{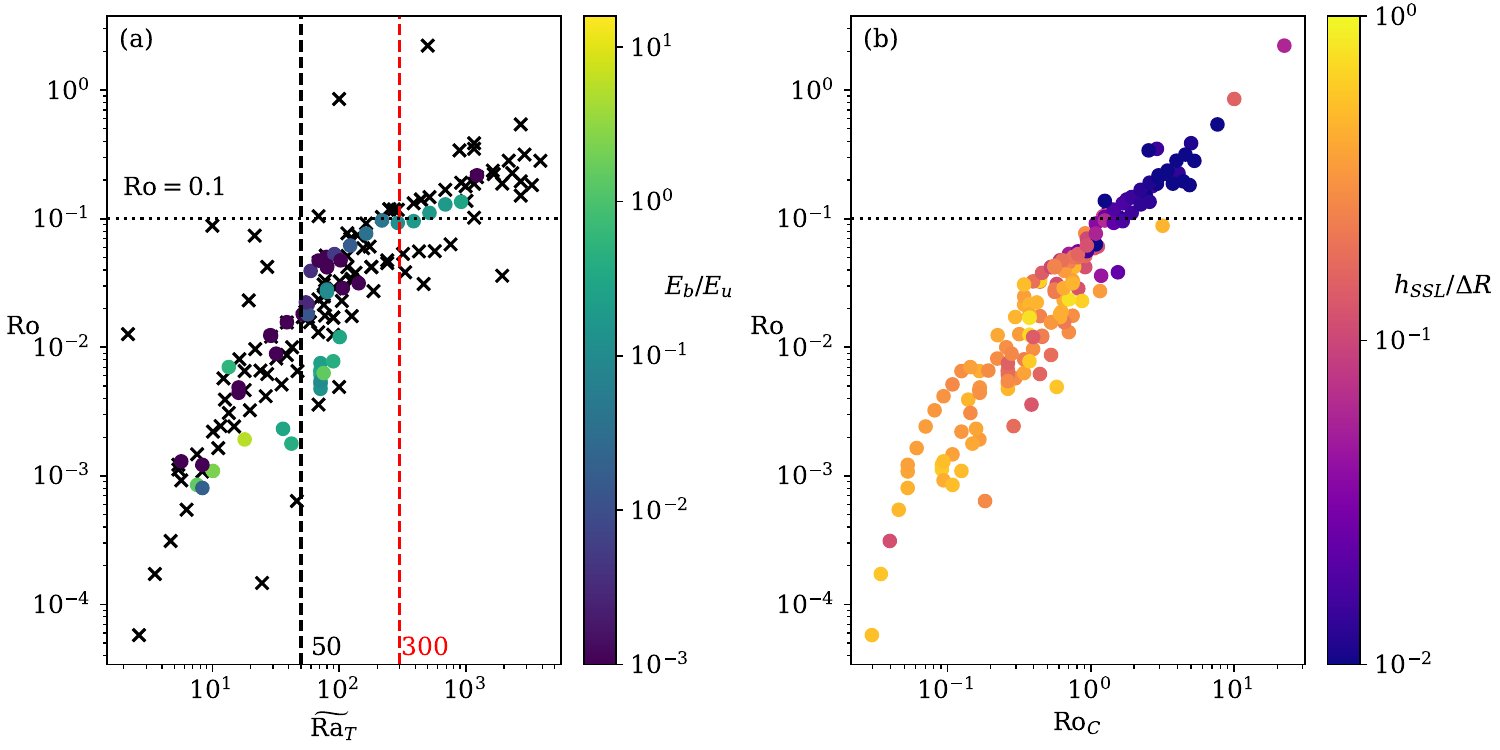}
    \caption{Rossby number $\Ro=\Ek\R$ plotted against (a) reduced Rayleigh number $\Rred$, coloured by the ratio of the magnetic to kinetic energies $E_b/E_u$, and (b) convective Rossby number $\Ro_C$, coloured by the size of the stably stratified layer. In (a), black crosses mark hydrodynamic simulations, while black and red dashed lines show the position of the transitions at $\Rred=50$ and $300$, respectively. Black dotted lines show $\Ro=0.1$; below this value the flow is considered to be rotationally constrained.}
    \label{fig:RoRB}
\end{figure}

Our simulations exist primarily in the weak field dynamo regime.
Figure~\ref{fig:RoRB}(a) shows the Rossby number $\Ro=\Ek\R$ against the reduced Rayleigh number $\Rred$, coloured by the ratio $E_b/E_u$ (with hydrodynamic simulations shown as black crosses). The maximum value of $E_b/E_u$ is $7.4$, with all but $4$ strong field simulations having $E_b/E_u<1$, and $60\%$ of the simulations having $E_b/E_u<0.1$. For reference, we expect a value in the range $10^{-4}<E_b/E_u<1$ for Jupiter \citep{pruzina2025planetary}. In all cases, the Rossby number is not significantly affected by the magnetic field, with the MHD simulations broadly collapsing onto the same curve as the hydrodynamics simulations. Figure~\ref{fig:RoRB}(b) shows $\Ro$ against the convective Rossby number $\Ro_C = \Ek\sqrt{\Rat/\Pr}$, coloured by $h_{SSL}$, which collapses better than in fig.~\ref{fig:RoRB}(a).  The rotationally constrained ($\Ro<0.1$) regime characteristic of planets is associated with a wide SSL (and small $\Rred$ and $\Ro_C$), although all simulations except one produced $\Ro<1$.

The majority of our simulations were carried out with $\Delta R=0.5$, $\Pr=0.3$ and $\Sc=3$. While we conducted a small number of simulations with different values of $\Delta R$ and $\Pr$, the Lewis number $\Le=\Sc/\Pr=10$ was kept constant. While this value is not far from the relevant value for gaseous planets \cite[e.g. $\Le \approx 30$ in Jupiter, see ][]{pruzina2025planetary}, we cannot exclude a Lewis number dependence in the curves seen in fig.~\ref{fig:RoRB}.
This moderate value of $\Le=10$ also somewhat limits the region of parameters where semi-convection can be studied (see equation \ref{eqn:semi-convectionrange}); a larger value of $\Le$ would increase the range of $R_\rho$ allowed in simulations from $1 \leq R_\rho \leq 3.25$ in this paper to $1 \leq R_\rho \lesssim 10$ for Jupiter.

\subsection{Dipolarity: effect of flow structure}
The magnetic fields of giant planets such as Jupiter and Saturn, as well as the Earth, are strongly dipolar, with Saturn's field also being very axisymmetric. As such, it is of considerable interest to identify trends in the form of the magnetic field generated by the semi-convective dynamo. We quantify the dipolarity and axisymmetry using the dipole and axisymmetry fractions
\begin{eqnarray}
	f_\text{dip} &=& \sqrt{\frac{\sum_{m=0}^1\left(2-\delta_{m0}\right)\left|\bm{b}\left(l=1,m,t\right)\right|^2}{\sum_{l\geq1,m\geq0}\left(2-\delta_{m0}\right)\left|\bm{b}\left(l,m,t\right)\right|^2}},\\
	f_\text{axi} &=& \sqrt{\frac{\sum_{l\geq0}\left(2-\delta_{m0}\right)\left|\bm{b}\left(l,m=0,t\right)\right|^2}{\sum_{l\geq1,m\geq0}\left(2-\delta_{m0}\right)\left|\bm{b}\left(l,m,t\right)\right|^2}},
\end{eqnarray}

where $l$ and $m$ are spherical harmonic degree and order, $\bm{b}(l,m,t)$ is the $(l,m)$ component of the magnetic field at the surface of the simulation $r=R_o$, and $\delta_{ij}$ is the Kronecker delta function. Note that these definitions are not totally standardised: it is common to define $f_\text{dip}$ using a spectrum truncated at the maximum $l$ of the current models of the target planet's magnetic field, while $f_\text{axi}$ is not a widely used measure.
\begin{figure}
    \centering
    \includegraphics[width=\textwidth]{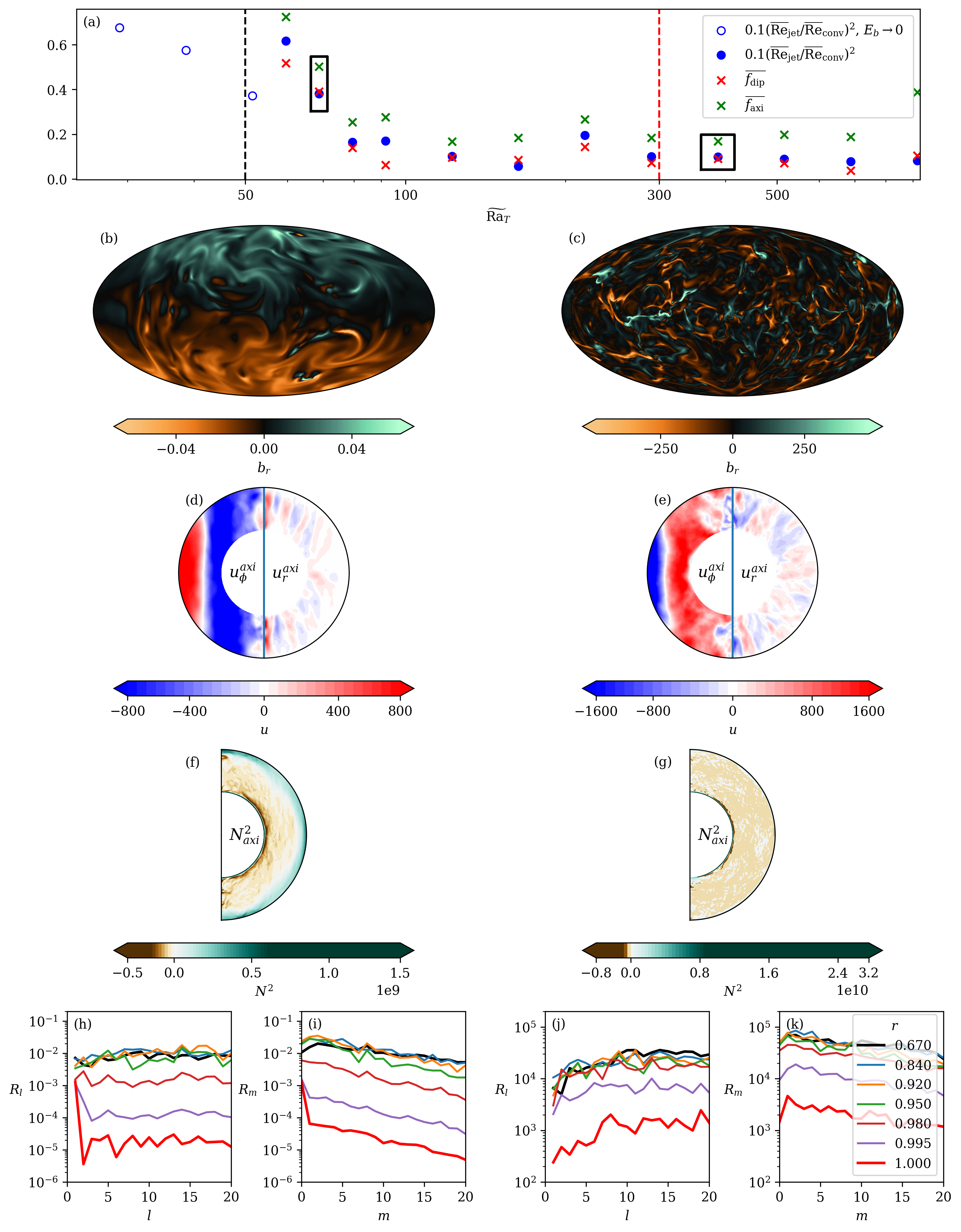}
    \caption{(a) Rescaled squared Reynolds number ratio $\left(\overline{\R_\text{jet}}/\overline{ \R_\text{conv}}\right)^2$, and dipolar and axisymmetric fractions $\overline{f_\text{dip}}$, $\overline{f_\text{axi}}$, along a transect of dynamo simulations with  $\Pm=1$, $\Pr=0.3$, $\Sc=3$, $\Delta R=0.5$, $\Ek=10^{-4}$, $R_\rho=1.2$ and $6.2\times10^6\geq\Rat\geq2.6\times10^8$. (b)--(g) Snapshots of the surface radial magnetic field $b_r$, axisymmetric part of the radial velocity field $u_r$, axisymmetric part of the azimuthal velocity $u_\phi$ and squared Brunt-V\"ais\"al\"a frequency $N^2$, for the two simulations indicated in black boxes in (a). (h)--(k) Magnetic energy spectra in spherical harmonic degree $l$ and order $m$ for each simulation, at a range of radii from the deep interior to the surface.}
    \label{fig:dipolar}
\end{figure}

Figure \ref{fig:dipolar}(a) shows how the dipole and axisymmetry fractions vary along a transect of dynamo simulations with $\Pm=1$, as well as the ratio of the squared jet/convective Reynolds numbers (representing the ratio of jet/convective kinetic energies). It appears that high values of $\overline{f_\text{dip}}$ and $\overline{f_\text{axi}}$ are associated with the jet-dominated regime. Figures \ref{fig:dipolar}(b)--(g) show snapshots of the magnetic field on the surface, the axisymmetric part of the velocity, and square Brunt-V\"ais\"al\"a frequency, for two simulations, one in the jet-dominated regime, and one where convection dominates. In the first simulation, the surface magnetic field is clearly strongly dipolar, with a clear partition between the two hemispheres. The velocity field is dominated by strong jets, with $u_r$ significantly smaller in magnitude than $u_\phi$. Figure~\ref{fig:dipolar}(f) shows a wide outer stably stratified layer. This simulation closely resembles that considered by \citet{pruzina2025planetary}. By contrast, in the second simulation,  the magnetic field (fig.~\ref{fig:dipolar}(c)) displays much more fine structure, with large contributions from higher-$l$ modes. The velocity jets are less columnar, and the radial velocity is closer in magnitude to the azimuthal velocity, showing a much higher proportion of energy in convection compared to jets than in fig.~\ref{fig:dipolar}(d). In fig.~\ref{fig:dipolar}(g), the convective zone takes up the entire domain, with much finer-scale density structures visible. 

Figure~\ref{fig:dipolar}(h)--(k) show the magnetic energy spectra at a range of radii for each simulation; in both cases, the spectrum is rather flat in the interior, and does not change significantly in the convective region. Further out, in the SSL, the spectrum decays with radius, with the dipole ($l=1$) component decaying at the slowest rate. In the jet-dominated regime, a wide SSL means that the field has decayed significantly and become strongly dipolar by the time it reaches the surface. By contrast, in the convection-dominated regime, the SSL is very thin, so the structure of the field at the surface is rather similar to that deep inside the shell. The effect of radius on the $m$-spectra is similar, with the $m>0$ modes decaying more rapidly than $m=0$ as radius increases, resulting in an axisymmetric-dominated field at the surface.

\subsubsection{A recipe for planetary magnetic fields?}
It has been proposed \citep[e.g.][]{stevenson1982reducing,cao2011saturn} that the magnetic field of Saturn is strongly axisymmetric and dipolar because of a stably stratified layer overlaying a convective zone where the field is generated. Simulations with an imposed SSL above the dynamo region confirm that this is possible \citep{yadav2022global}. Our simulations of semi-convection show that this structure can be generated spontaneously by semi-convection. With this in mind, we can identify the correct regions of parameter space to produce planet-like magnetic fields. A thick stably stratified layer is needed to filter the field, so a small-as-possible value of $\Rred$ is preferable. However, velocities must also be large enough to generate a dynamo, so $\Rred$ must also be large enough to provide these turbulent flows. As such, a value $\Rred$ just greater than $50$ should be ideal. Figure~\ref{fig:innerlayerwidth} showed that the scaling for $h/\Delta R$ did not depend significantly on domain depth; therefore, a larger SSL can also be obtained by simulating a thicker shell.

\begin{figure}
    \centering
    \includegraphics[width=\textwidth]{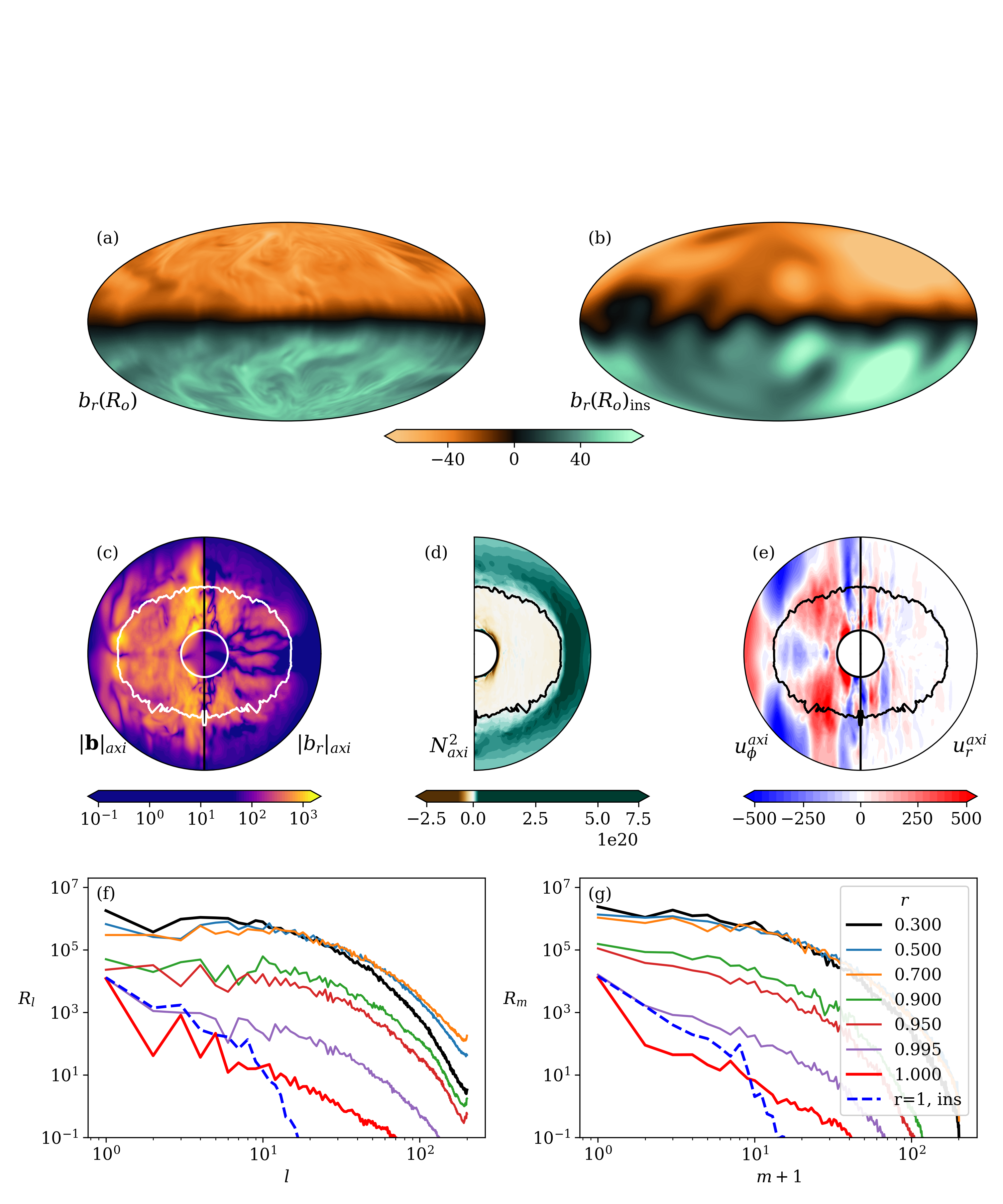}
    \caption{Results from a simulation with $\Ek=10^{-5}$, $\Rac=4.2\times10^9$, $\Rat=3.5\times10^8$, $\Pr=0.3$, $\Sc=3$, $\Pm=1$,  $\Delta R=0.8$, showing (a) surface magnetic field $b_r$, (b) surface field when the fluid in $0.7\leq r \leq 1$ is insulating,  and axisymmetric parts of the (c) magnitude of the magnetic field $|\bm{b}|$, and radial magnetic field $|b_r|$ (d) square of the Brunt-V\"ais\"al\"a frequency $N^2$, (e) radial and azimuthal velocity, (f) $l$-spectrum and (g) $m$-spectrum of the magnetic energy at a range of radii, with the surface spectra for the insulating outer layer case shown as blue dashed lines. The boundary between convective and stably stratified regions, and the inner boundary of the fluid domain, are marked on the polar plots in white/black.}
    \label{fig:widedipaxi}
\end{figure}

 Based on these criteria, we present the results of a simulation for $\Ek=10^{-5}$, $\Rac=4.2\times10^9$, $\Rat=3.5\times10^8$, $\Pr=0.3$, $\Sc=3$, $\Pm=1$, in a shell with $\Delta R=0.8$.  Note that these parameter values are rather moderate, so the simulation is not particularly computationally challenging, with a resolution of only $(N_R,N_l,N_m)=(256,200,200)$ in the radial direction and spherical harmonic $l$ and $m$. These values are chosen such that $\Rred=75$; just beyond the transition. Figure~\ref{fig:widedipaxi}(a) shows the surface radial magnetic field for this simulation; the field is very strongly dipolar and axisymmetric. This field is also rather strong, compared to the majority of our simulations, with an energy ratio of $E_b/E_u = 1.52$. For comparison, fig.~\ref{fig:widedipaxi}(b) shows the surface field if the fluid in the layer $r_L\leq r\leq 1$ is insulating (calculated by setting the toroidal component of $\bm{b}$ to zero for $r>r_L$, and the imposing poloidal component $P(r) = P(r_L) \times (r/r_L)^{-l-1}$, with $r_L=0.7$). This `truncated'  field is noticeably less dipolar and axisymmetric than the full field, and in general shows larger-scale structures.
 
 Figure~\ref{fig:widedipaxi}(c)--(e) show the axisymmetric ($m=0$) part of the magnitude of the total and radial magnetic fields, squared Brunt-V\"ais\"al\"a frequency, and azimuthal and radial velocity fields. The black line shows the boundary between the convective zone and SSL. Inside the convective zone, the radial and azimuthal velocities are of a similar magnitude, and there is a strong magnetic field; in the SSL, the radial velocity becomes small, and the strength of the magnetic field drops (particularly the radial component).

 Figure~\ref{fig:widedipaxi}(f)--(g) shows spectra of the magnetic energy at a range of radii; deep in the interior, the $l$ and $m$ spectra are both rather flat, with no variation with depth. Closer to the surface, the power drops off sharply with radius, resulting in a surface field with very strongly dominant dipole ($l=1$) and axisymmetric ($m=0$) components, with $\overline{f_\text{dip}}=0.95$ and $\overline{f_\text{axi}}=0.99$. Saturn's magnetic field has dipole and axisymmetric fractions of $f_\text{dip}\approx0.99$ and $f_\text{axi}\approx0.985$ \citep{cao2020landscape}, so our simulation compares rather favourably. The blue dashed lines in figs.~\ref{fig:widedipaxi}(f)--(g) show the surface power spectra of the truncated magnetic field (with an insulator for $r_L\leq r\leq 1$). Compared to the full spectra, these profiles show a slower decay at low values of $l$ and $m$, with values of $\overline{f_\text{dip}}=0.87$ and $\overline{f_\text{axi}}=0.92$ --- markedly smaller than for the full magnetic field. This greater dominance of $l=1$ and $m=0$ of the full field demonstrates that the flow in the SSL actively filters the magnetic field, rather than simply acting as an insulating layer where the field decays \citep[as discussed by, e.g.][]{stevenson1982reducing}. 

\section{Conclusions}\label{sec:conclusions}
In recent years, significant attention has been focused on double-diffusive convection in astrophysical bodies, with new structure models of gas giants proposing that large regions of planetary cores may be unstable to semi-convection. With this in mind, we have presented a detailed numerical study of semi-convection in rotating spherical shells, in the low-$\Pr$ regime appropriate for these astrophysical applications, with the additional motivation of producing a planetary-like dynamo.

\subsection{Results summary}

\begin{figure}
	\includegraphics[width=\textwidth]{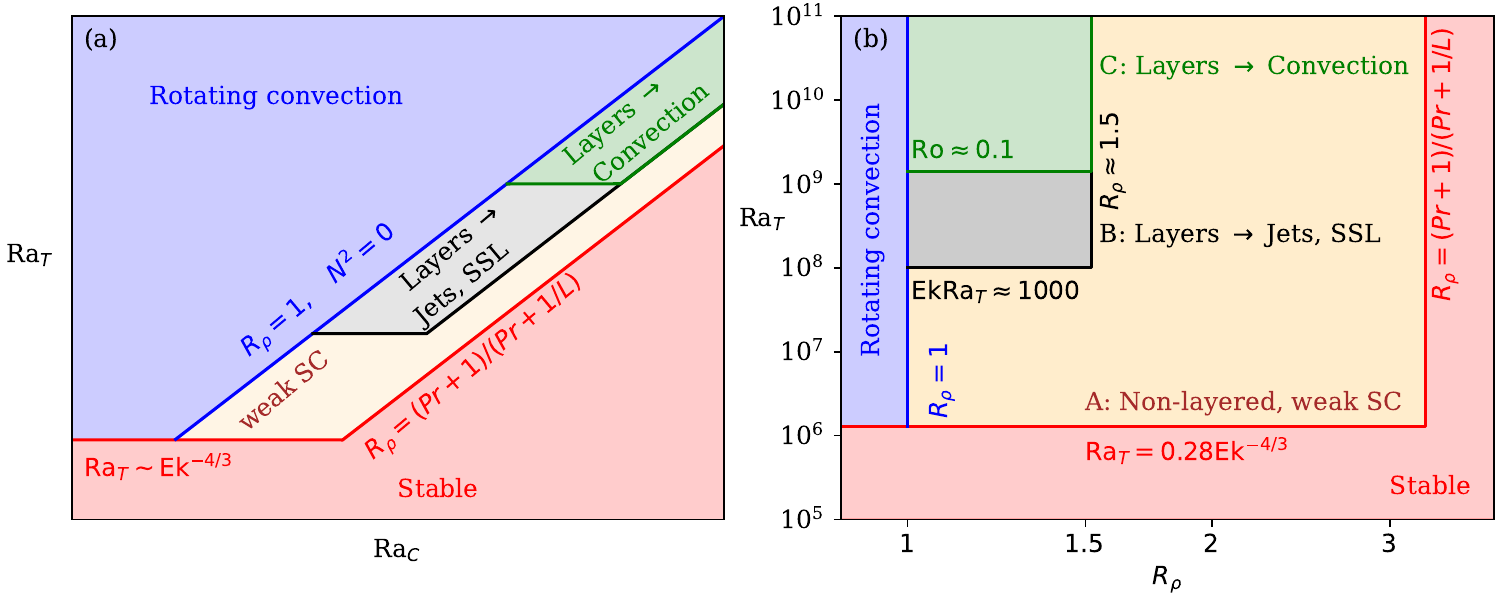}
	\caption{Regime diagram showing the variety of flows observed in rotating semi-convection in spherical shells: (a) General sketch in $\Rac$-$\Rat$ space; (b) for $\Ek^{-5}$, $\Pr=0.3$, $\Sc=3$ in $\Rat$ -- $R_\rho$ space. The coloured lines and areas are consistent between the two panels. In the grey and green regions, the flow first develops density layers, followed by either zonal jets and a wide SSL, or a convective region spanning the domain.}
	\label{fig:regimediagram}
\end{figure}

\begin{landscape}
	\begin{figure}
		\includegraphics[width=1.6\textwidth]{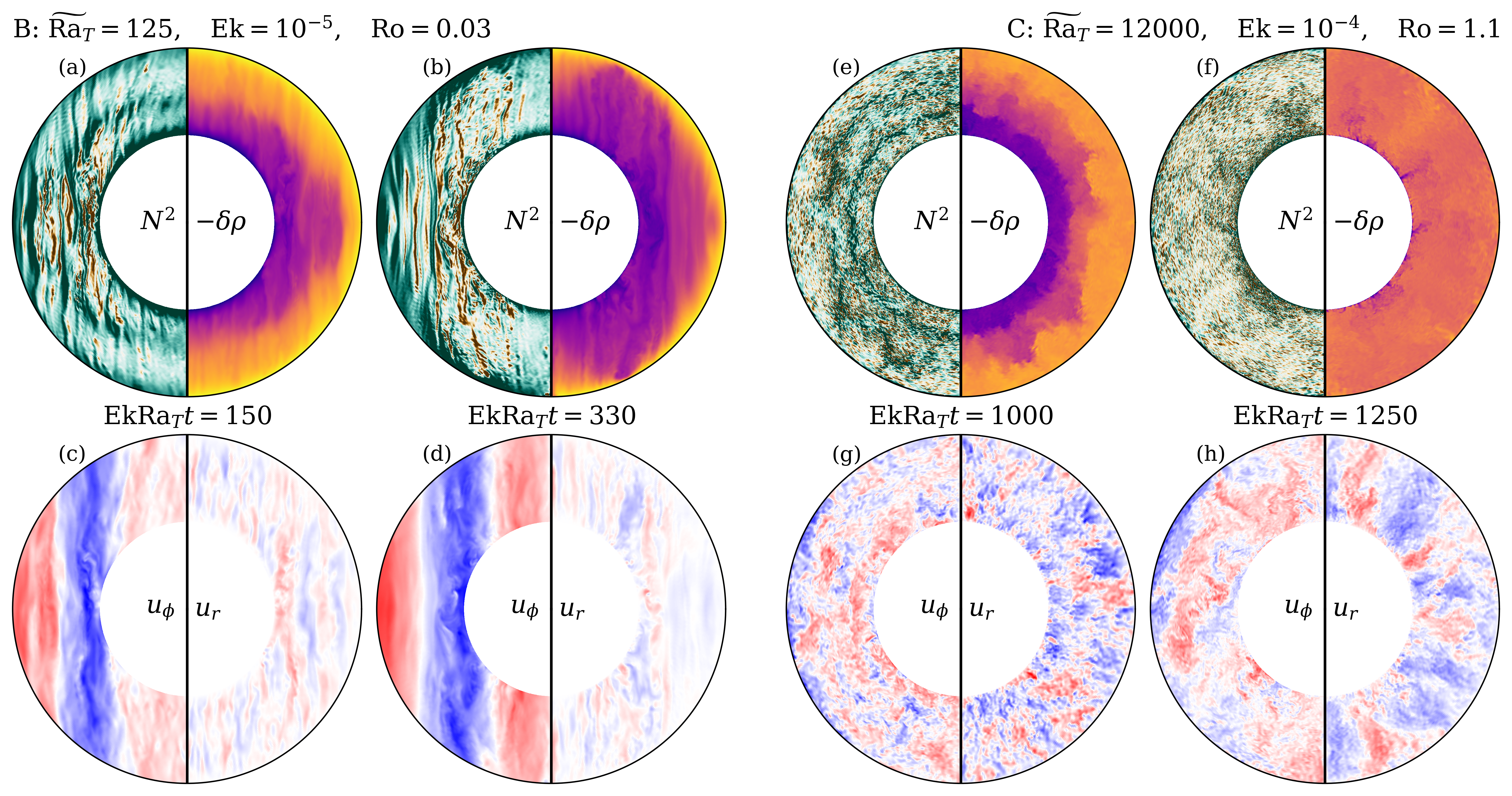}
		\caption{Examples of the two main flow regimes seen in our simulations, showing the density anomaly $-\delta\rho=(T+T_0)-(C+C_0)$, Brunt-V\"ais\"al\"a frequency $N^2$, and radial and azimuthal velocities, in meridional slices. Each simulation is shown at two times; one in the layered phase, and one in the final state. Panels (a)--(d) show a simulation for $\Rred=125$ that forms two layers, and finishes in the jet-dominated regime; (e)--(h) show a three layer simulation with $\Rred=12000$ that develops into the convection-dominated regime.}
		\label{fig:megaplot}
	\end{figure}
\end{landscape}

We have performed a large number of numerical simulations of semi-convection using the XSHELLS spherical harmonic solver, for several Ekman numbers and a wide range of thermal and compositional Rayleigh numbers. We summarise the different varieties of flow in the regime diagram in fig.~\ref{fig:regimediagram}.  High resolution plots characteristic of the two main flow regimes of interest are shown in fig.~\ref{fig:megaplot}.

In general, we considered simulations along transects with a constant density ratio $R_\rho=1.2$, passing through the `weak SC', `Jets, SSL', and `Convection' regimes seen in fig.~\ref{fig:regimediagram}. Note that, for constant $R_\rho$, increasing the thermal forcing $\Rat$ also increases the compositional stratification $\Rac$ at the same rate. We also considered limited number of other simulations with $\Rac=5\times10^9$ fixed, and $1\leq R_\rho\leq3.3$.

Past the onset of rotating semi-convection, for weak thermal forcing (region A in fig.~\ref{fig:regimediagram}(b)), the semi-convection instability grows, eventually saturating in a weakly non-linear state, with transport happening almost entirely by diffusion. When the forcing is stronger, however, complex non-linear behaviour is seen (regions B and C). After an exponential growth phase, well-mixed layers of relatively constant density develop, separated by interfaces with large density gradients. Layering follows the widely accepted $\gamma$-instability, which states that layers will develop if the ratio of the thermal to compositional fluxes decreases as a function of the ratio of their gradients \citep{radko2003mechanism,rosenblum2011turbulent,mirouh2012new}. However, the scale on which these layers form scales with $(\Ek\Rat)^{-1/3}$, so if the thermal forcing is not sufficiently strong, the domain is not large enough for layers to form. This layer size contrasts with the scaling $\Rat^{-1/4}$ for non-rotating semi-convection \citep{mirouh2012new}. 

The layers merge over a timescale $\sim(\Ek\Rat)^{-1}$, via a pattern where relatively weak interfaces grow progressively weaker, while strong interfaces intensify  \cite[the B-merger of][]{radko2007mechanics}. For strong forcing and weak influence of rotation (region C in fig.~\ref{fig:regimediagram}(b), with an example shown in fig.~\ref{fig:megaplot}(f),(h)), the layers eventually merge to a single convective region occupying almost the entire domain, with only narrow stably stratified boundary layers at the top and bottom.  However, for weaker forcing and more dominant rotation (region B, fig.~\ref{fig:megaplot}(e),(g)), a remnant of the layers remains, with an inner convective regione overlain by a stably stratified layer. In this case, the flow is dominated by strong zonal jets, with convective motions being relatively weaker.

The transition between these two regimes is controlled by the Rossby number.
When $\Ro\lesssim0.03$, the stably stratified layer is identified as a thermal boundary layer influenced by rotation, with its height $h_{SSL} \sim \Ek^{-4/9}\Rat^{-1/3}$,
resulting in the wide SSL seen in fig. 15(b). Where $\Ro\gtrsim0.1$, there is instead a non-rotating boundary layer scaling $h_{SSL}\sim\Rat^{-1/3}$. This much narrower boundary layer results in the very wide convective zone in fig. 15(f). In between these two regimes, there
is a continuous transition, where $h_{SSL}$ varies quickly.
 We note that, while the initial layered phase is only a short-term transient, it is essential for the development of large scale density anomalies. Where layering does not occur (either because there is no $\gamma$-instability, or because the domain is not large enough for layers to form), only weak flows develop, and the density perturbation remains very small (region A in fig.~\ref{fig:regimediagram}).

This distinction between convective and SSL regimes has not been documented in previous studies, which have primarily focused on a periodic box geometry. In rotationally constrained spherical flows, the tendency to form columnar structures is in competition with the radial merging and mixing behaviour of semi-convection, resulting in the jet-dominated regime where the SSL is significant.

In general, the flow strength (measured by the Reynolds number) increases with forcing (measured by $\Rred$). When the SSL is large, the flow is dominated by zonal jets in the SSL, with convective motions confined to the inner convective region. As the SSL reduces in size, the relative strength of zonal jets reduces, with convective motions becoming dominant at the $h_{SSL}\sim(\Ek/\Pr)^{1/2}$ transition.
All our simulations collapse onto a single master curve for $\Ro$ in terms of the convective Rossby number $\sqrt{\Rat\Ek^2/\Pr}$, with rotationally constrained flows ($\Ro<0.1$) corresponding to the jet-dominated, wide-SSL regime. 

Magnetohydrodynamic simulations reveal that the flow regime (jet or convection-dominated) has a significant effect on the magnetic field.
In the jet-dominated regime, the dynamo exists in the deep convection region, with the magnetic field smoothed out within the SSL, resulting in a weak surface field dominated by the dipolar, axisymmetric components.
This smoothing in the SSL is more efficient than a simple insulating layer, resulting in stronger dipole and axisymmetry fractions, which implies that strong zonal flows in the SSL actively filter the field to produce a strong axial dipole dominance. In convection-dominated simulations, the full multipolar field survives to the surface, with a relatively flat spectrum barely altered by a very thin stably stratified thermal boundary layer. 

Stably stratified layers are thus good candidates to produce axisymmetric magnetic field similar to the one of Saturn. When the SSL is large enough to filter the magnetic field, while still keeping vigorous convection below, the system generates an axisymmetric planetary-like magnetic field. We note however, that even without the SSL, it is possible to generate a planetary-like magnetic field in a thin convective layer \citep[e.g.][]{schaeffer2025energetically}.

 This mechanism of a deep SSL filtering the magnetic field to result in the observed dipolar/axisymmetric field has been theorised for several decades \citep[e.g.][]{stevenson1982reducing}, and in recent years simulations with an imposed SSL above a convective zone have produced results that compare very favourably to the magnetic field of Saturn \citep[][]{yadav2022global}. Our work demonstrates that both the convective layer generating the field, and the SSL to filter it, can be the result of the self-organisation of the fluid, in a region with an initially semi-convective density profile. The dipole and axisymmetry fractions of the magnetic field in our simulation compare favourably to those observed in Saturn \citep{cao2020landscape}. We do not consider other metrics, such as flow structure or dipole tilt, as we do not claim to present a planetary model in this study, but rather to demonstrate the possible applications of a dynamo generated by semi-convection layering, and spark interest for future studies.

\subsection{Perspectives}

To comment on planetary applications of this work, it is necessary to consider the vastly different parameters in real planets compared to our simulations. For Jupiter,  \citet{pruzina2025planetary} gave the estimates $\Pr\approx 0.07$, $\Sc\approx2$ and $\Ek=O(10^{-19})$. \citet{mirouh2012new} found that increasing $\Le$ and decreasing $\Pr$ both increased the value of $R_{\rho,layer}$, resulting in significantly larger regions B and C in fig~\ref{fig:regimediagram}(b) .  Likewise, the much smaller value of $\Ek$ in planets means that the bottom boundary of the layering region (marked $\Ek\Rat=1000$) will be below the onset of instability itself, so layering is always expected in regions with appropriate values of $R_\rho$. Hence, for planetary parameters, it appears highly likely that the behaviour in regime B is relevant. With the characterisation of the SSL in regime B as the maximum size of a diffusive layer in rotating convection, it is possible that measurements of the stable layer size could contribute to constraining the heat flux in planets.

In this work, we have made a number of assumptions that should be discussed. Firstly, we considered the Boussinesq equations, neglecting differences in density except in conjunction with gravitational acceleration.  For planetary applications, the validity of this approximation hinges on density variations being fairly small across the semi-convection region. In reality, the density is expected to vary rather significantly, especially for larger semi-convection regions as proposed for Saturn. However, for an initial survey, and from hydrodynamic interest, the Boussinesq approximation reveals a lot about the dynamics of semi-convection. Being simpler computationally than alternatives (such as the anelastic approximation), it allows the investigation of more extreme parameter values. 

We adopted constant-value boundary conditions on $T$ and $C$ to approximate the effect of convective regions above and below the semi-convective layer, which may not be strictly accurate (particularly for Saturn, where a convective zone is not necessarily expected for $r<R_i$). In the layered phase of evolution, our simulations are subject to the $\gamma$-instability followed by layer mergers, similarly to periodic box simulations of semi-convection. The same behaviour has also been documented recently with fixed-flux conditions in cartesian boxes \citep{tulekeyev2024constraints} and full spheres \citep{fuentes20253d}, suggesting that these layering behaviours are generic. For comparison, we have run a small number of preliminary simulations with fixed-flux conditions, and find that the transition between jet-dominated and convection-dominated regimes appears to still be present; further study will be required to say more about the effect of the boundary conditions.

We also assumed that both gravitational acceleration and electrical conductivity (influencing the magnetic diffusivity) were constant with depth. In planetary semi-convection, neither of these assumptions is strictly true, and allowing for a variable conductivity would be an interesting and useful extension to this work. Similarly, allowing for the density ratio $R_\rho$ to vary in space would provide a more realistic planetary simulation.

Several planetary models explain the decrease with radius of helium concentration as `helium rain' \citep[e.g.][]{howard2024evolution}. With a saturation concentration $C_s(T)$ defined as a function of temperature, any composition exceeding this value $C>C_s$ is considered to condense, leading to a `moist convection' model. \citet{leconte2017condensation} show that, in the limit of fast condensation, semi-convection is completely suppressed, causing no fluid motion. They suggest that gas giants are indeed in this limit, which would imply that in the helium rain layer, heat transfer occurs only through radiation. However, if condensation is not totally instantaneous, semi-convection may not be fully inhibited, and the inclusion of a semi-convection region can explain all the observed gravitational and atmospheric constraints for Jupiter and Saturn \citep{leconte2012new}. A further extension could be made to this work by the inclusion of condensation, as a third process competing against semi-convection and rotation.

\begin{acknowledgments}
\textit{Acknowledgements:} The authors thank F. Debras for discussions about the existence of semi-convection layers in gaseous planets, and their potential to drive a dynamo. We also thank J. Vidal for assistance with the SINGE code. We additionally thank three anonymous referees for their help to improve the manuscript.

\textit{Funding:} Work funded by the ERC under the European Union’s Horizon 2020 research and innovation program via the THEIA project (grant agreement no. 847433), by the European High-Performance Computing Joint Undertaking under grant agreement 101093038 (ChEESE-2P), and by the NumPEx PEPR program
ANR-22-EXNU-0006.
Computations on the GRICAD infrastructure (https://gricad.univ-grenoble-alpes.fr), and HPC resources (Jean Zay V100 and H100) of IDRIS under allocation AD010413621 and A0160407382 attributed by GENCI (Grand Equipement National de Calcul Intensif).

\textit{Author contribution statement:} PP produced the simulations, the figures, and the draft. DC came up with the initial research concept, and obtained the funding. NS modified the numerical code and helped with computations. DC and NS obtained computation hours from GENCI. All authors discussed each step of the work and the results; all edited the final manuscript.

\textit{Declaration of interests:} The authors report no conflicts of interest.

\textit{Data availability statement:} Summary data of all the simulations are available at  doi:10.5281/zenodo.18314793
 
\end{acknowledgments}

\appendix
\section{Linear stability of semi-convection in a rotating spherical shell}\label{sec:linearonset}
In this appendix, we report on the linear onset of instability, determining the shape of the unstable regions in Rayleigh number space and the dependence on the Ekman number. 
For three values of $\Ek$, we find the onset of instability for a wide range of values of $\Rac$. We first show results with fixed-value boundary conditions \eqref{eqn:bcs} on $T$ and $C$, before briefly discussing the effect of fixed-flux conditions.

We use the SINGE linear eigensolver \citep{vidal2015,monville2019rotating} to find the onset of instability. At each value of $\Rac$, and each spherical harmonic order $m$, SINGE finds the critical value $\Rat^c(\Rac,m)$ such that the growth rate of the eigenmode is 
\begin{equation}
	\sigma(\Rat^c,\Rac,m)=0.
\end{equation}
We define the critical Rayleigh number as
\begin{equation}
	\RaTc(\Rac) = \text{min}_{m}\{\Rat^c(\Rac,m)\}.
\end{equation}

For thermal convection (i.e. $\Rac=0$) in a rotating spherical shell, \citet{dormy2004onset} found that, in the limit of low $\Ek$, the onset of instability followed the scaling $\RaTc\sim \mathcal{R}_0\Ek^{-4/3} + \mathcal{R}_1\Ek^{-1}$ for a relatively thin shell ($\Delta R = 0.35$). We begin by confirming that this extends to our wider shell by using SINGE to find the critical value $\RaTc(0)$ for three values of $\Ek$. These values are reported in Table~\ref{tab:convectiononset}, and fit the scaling of \citet{dormy2004onset} perfectly.
\begin{table}
	\centering
	\begin{tabular}{c|c|c|c}
		$\Ek$&$\RaTc$&$m_\text{crit}$&$\omega_\text{crit}$\\
		$10^{-4}$& $3.45\times10^5$&$8$&$-5.93\times10^2$ \\
		$10^{-5}$& $4.24\times10^6$&$16$ &$-2.84\times10^3$\\
		$10^{-6}$& $5.74\times10^7$& $28$&$-1.89\times10^4$
	\end{tabular}
	\caption{Critical Rayleigh number and spherical harmonic order at the onset of rotating thermal convection $\Rat^{rot}=\RaTc(\Rac=0)$ for three values of $\Ek$. These values follow the scaling $\Rat^{rot}\sim 0.28\Ek^{-4/3} + 30\Ek^{-1}$, matching  the theory of \citet{dormy2004onset}. For completeness, we also give the drift rate $\omega_{crit}$, in units of viscous time}
	\label{tab:convectiononset}
\end{table}

\begin{figure}
	\centering
	\includegraphics[width=\textwidth]{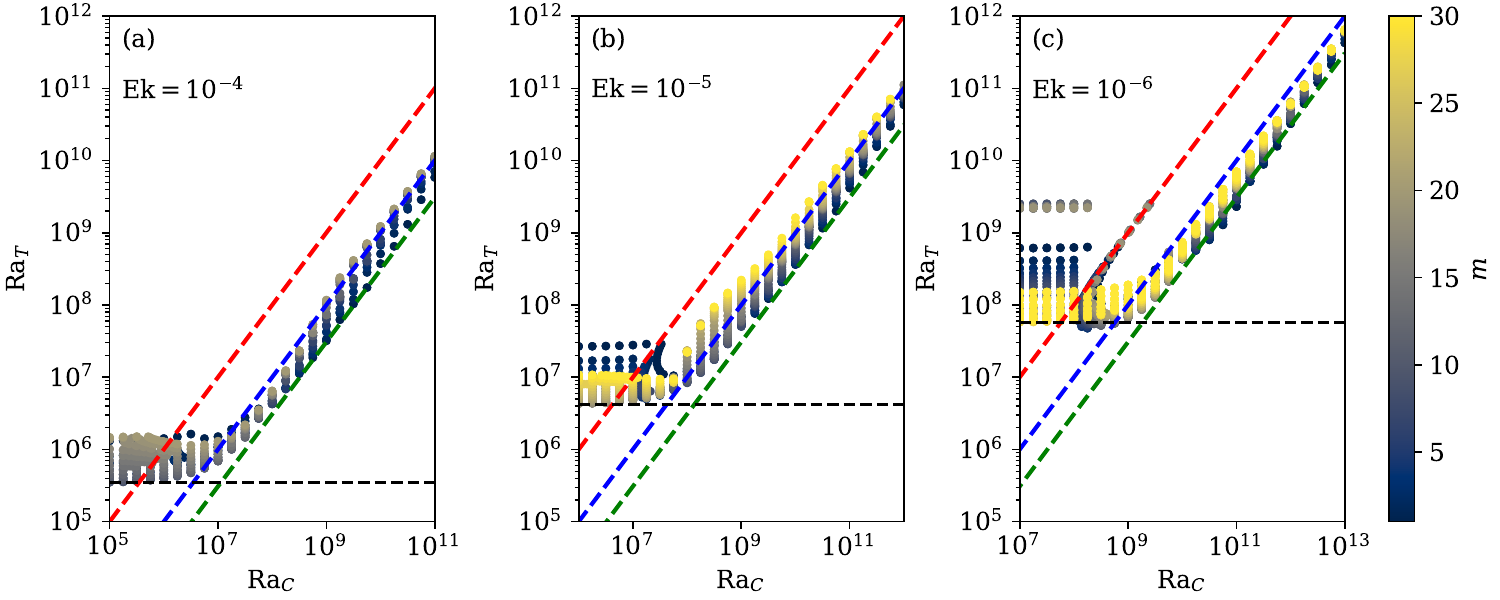}
	\caption{Linear onset of semi-convection computed with SINGE, for $\Delta R=0.5$, $\Pr=0.3$, $\Sc=3$, with three values of $\Ek$ and spherical harmonic order $1\leq m\leq 40$. Each dot shows the minimum $\Rat$ for instability, at that choice of $\Rac$ and $m$. Diagonal lines show (red) $\Rat=\Rac$, (blue) $\Rat=\Rac/\Le$, and (green) the limit of non-rotating semi-convection given by $\Rat^{SC}$, and horizontal black lines show the limit of rotating convection reported in Table~\ref{tab:convectiononset}.}
	\label{fig:singeonset}
\end{figure}

Figure~\ref{fig:singeonset} shows the onset of instability in Rayleigh-number space, for three values of $\Ek$. For a range of values of $\Rac$, $\Rat^c(\Rac,m)$ is plotted, i.e. the minimum value of $\Rat$ for instability at that value of $\Rac$ and $m$, coloured by the value of $m$. For small values of $\Rat$, the system is stable; as $\Rat$ increases, the critical threshold $\RaTc\left(\Rac\right)$ is passed and the system becomes unstable. For each value of $\Ek$, there are two limiting regimes for the global onset; that of rotating convection, where $\RaTc\sim\Rat^{rot}\sim\Ek^{4/3}$ (cf. Table~\ref{tab:convectiononset}), and that of non-rotating semi-convection, where $\RaTc\sim\Rat^{SC}$ (see \eqref{eqn:semiconvRa}). Except for a transition region where the two are of similar size, $\RaTc\sim\text{max}\{\Rat^{rot},\Rat^{SC}\}$. The onset of instability is at small scale (large $m$) for $\Rac\lesssim\Rat^{rot}$, and large scale ($m=1$) for $\Rac\gtrsim\Rat^{rot}$

For each individual value of $m$, the onset curve follows a similar overall trend to the global onset, with $\Rat^c(m)$ tending to a constant value at low $\Rac$, and parallel to $\Rat^{SC}$ at high $\Rat$. However, in the transition region near $\Rac\approx\Rat^{rot}$, the behaviour is more complex. Large values of $m$ show a smooth transition between the two limits. For smaller $m$, the onset curve forms a `tongue' for $\Rac/\Le\leq\Rat\leq\Rac$, extending the unstable range below the small $\Rac$ limit for that value of $m$. For $\Ek=10^{-4}$ and $10^{-5}$, this tongue does not affect the overall region of instability. However, for $\Ek=10^{-6}$, at the smallest values of $m$, the tongue reaches values of $\Rat$ beneath the rotating convection limit, slightly extending the unstable region 

The existence of this tongue, where, as the stabilising gradient $\Rac$ is increased, the forcing required for instability $\RaTc$ decreases, fits with the findings of \citet{busse2002low} and \citet{monville2019rotating}. They found that the tongue extended the range of instabilities for all values of $\Ek$ tested, rather than only the smaller ones. However, \citet{monville2019rotating} considered fixed-flux boundary conditions on $T$ and $C$, compared to our fixed-value conditions. 

\begin{figure}
	\includegraphics[width=\textwidth]{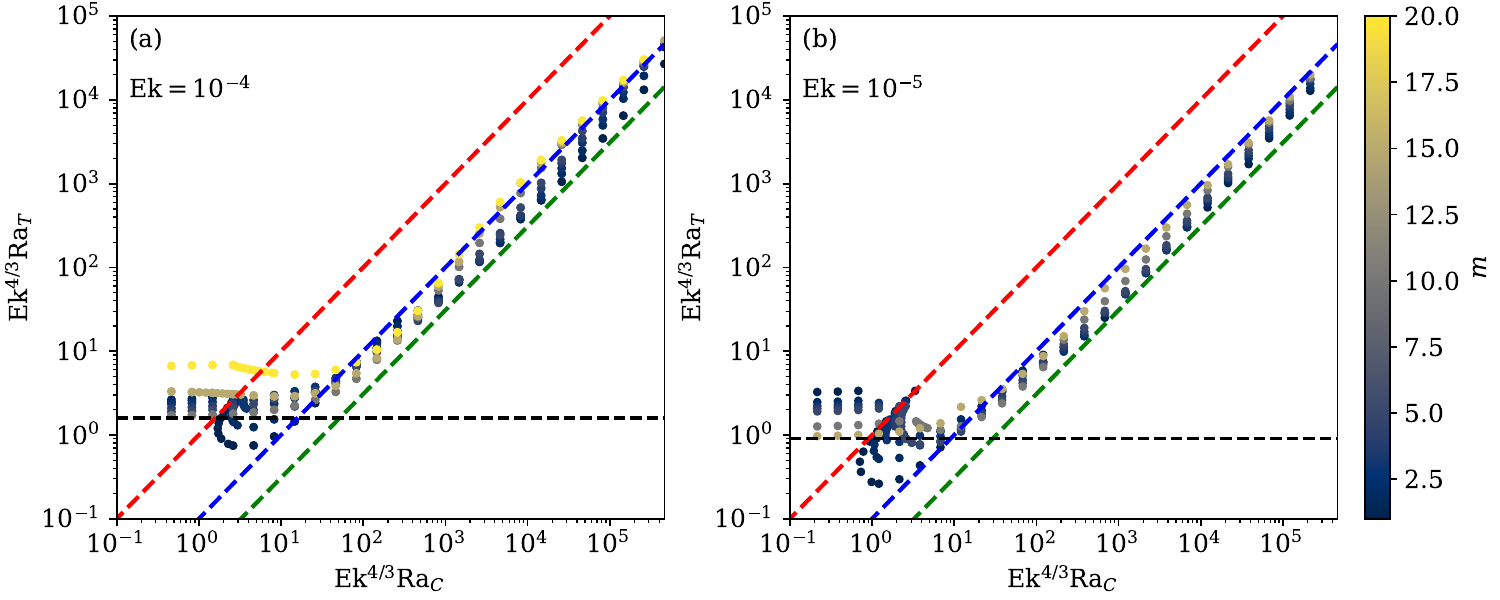}
	\caption{Linear onset of rotating semi-convection with fixed temperature and salinity flux boundary conditions, for $\Delta R = 0.5$, $\Pr=0.3$, $\Sc=3$, with two values of $\Ek$ and spherical harmonic orders $0\leq m\leq 20$. Diagonal lines show (red) $\Rat=\Rac$, (blue) $\Rat=\Rac/\Le$, and (green) the limit of rotating semi-convection given by $\Rat^{SC}$. Horizontal black dashed lines show the limit of rotating thermal convection (at $\Rac=0$).}
	\label{fig:singeflux}
\end{figure}

Figure~\ref{fig:singeflux} shows the onset of rotating semi-convection, for fixed-flux boundary conditions,
\begin{equation}
	\frac{\partial T}{\partial r} = \frac{\partial C}{\partial r} = 0 \text{ at } r=1-\Delta R, 1,
\end{equation} 
and a reduced selection of values of $m$ compared with fig.~\ref{fig:singeonset}. We see that for both $\Ek=10^{-4}$ and $10^{-5}$, the tongue reported by \citet{monville2019rotating} for the fingering regime is present, with $\RaTc$ reduced significantly below the onset of rotating convection. This contrasts with fig.~\ref{fig:singeonset}, where the tongue only extended $\RaTc$ beneath $\Rat^{rot}$ for $\Ek=10^{-6}$.

\bibliographystyle{jfm}
\bibliography{bibliography}

\end{document}